\documentclass[fleqn,10pt,twocolumn]{wlscirep}
\usepackage[utf8]{inputenc}
\usepackage[T1]{fontenc}
\usepackage{subfig}
\usepackage{caption}
\usepackage{mwe} 
\usepackage{amsmath}
\usepackage{soul}
\usepackage{amssymb}
\usepackage{bm}
\usepackage{ulem}
\usepackage{makecell}
\usepackage[version=4]{mhchem}
\usepackage{multirow}
\usepackage{tikz}

\usepackage[page]{appendix}

\let\oldAA\AA
\renewcommand{\AA}{\text{\normalfont\oldAA}}

\newcommand{\fig}{Fig.}
\newcommand{\Fig}{Figure}
\newcommand{\figref}[1]{\fig~\ref{#1}}

\newcommand{\Figref}[1]{\Fig~\ref{#1}}

\newcommand{\figvref}[1]{\fig~\vref{#1}}
\newcommand{\Figvref}[1]{\Fig~\vref{#1}}

\newcommand{\tabref}[1]{Table~\ref{#1}}
\newcommand{\Tabref}[1]{Table~\ref{#1}}
\newcommand{\tabvref}[1]{table~\vref{#1}}
\newcommand{\Tabvref}[1]{Table~\vref{#1}}

\renewcommand{\eqref}[1]{Eq.~(\ref{#1})}
\newcommand{\Eqref}[1]{Equation~(\ref{#1})}
\newcommand{\eqvref}[1]{equation~(\ref{#1}) \vpageref{#1}}
\newcommand{\Eqvref}[1]{Equation~(\ref{#1}) \vpageref{#1}}

\newcommand{\secref}[1]{Section~\ref{#1}}
\newcommand{\Secref}[1]{Section~\ref{#1}}
\newcommand{\secvref}[1]{Section~\ref{#1} \vpageref{#1}}
\newcommand{\Secvref}[1]{Section~\ref{#1} \vpageref{#1}}

\definecolor{myblue}{HTML}{1f77b4}
\definecolor{myorange}{HTML}{ff7f0e}
\definecolor{mygreen}{HTML}{2ca02c}
\definecolor{myred}{HTML}{d62728}
\definecolor{cream}{RGB}{222,217,201}

\graphicspath{{figures/}{figures/}}

\makeatletter
\def\input@path{{figures/}}
\makeatother

\usepackage[switch]{lineno} 
\title{Performance of universal machine-learned potentials with explicit long-range interactions in biomolecular simulations}

\author[1, $^\ast$]{Viktor Zaverkin}
\author[2]{Matheus Ferraz}
\author[1]{Francesco Alesiani}
\author[1, 3]{Mathias Niepert}

\affil[1]{NEC Laboratories Europe GmbH, Kurf{\"u}rsten-Anlage 36, 69115 Heidelberg, Germany}
\affil[2]{NEC OncoImmunity AS, Forskningsparken, Gaustadall{\'e}en 21, 0349 Oslo, Norway}
\affil[3]{Institute for Artificial Intelligence, University of Stuttgart, Universit{\"a}tsstra{\ss}e 32, 70569 Stuttgart, Germany}
\affil[$\ast$]{viktor.zaverkin@neclab.eu}

\begin{abstract}
	Universal machine-learned potentials promise transferable accuracy across compositional and vibrational degrees of freedom, yet their application to biomolecular simulations remains underexplored. This work systematically evaluates equivariant message-passing architectures trained on the \mbox{SPICE-v2} dataset with and without explicit long-range dispersion and electrostatics. We assess the impact of model size, training data composition, and electrostatic treatment across in- and out-of-distribution benchmark datasets, as well as molecular simulations of bulk liquid water, aqueous \ce{NaCl} solutions, and biomolecules, including alanine tripeptide, the mini-protein Trp-cage, and Crambin. While larger models improve accuracy on benchmark datasets, this trend does not consistently extend to properties obtained from simulations. Predicted properties also depend on the composition of the training dataset. Long-range electrostatics show no systematic impact across systems. However, for Trp-cage, their inclusion yields increased conformational variability. Our results suggest that imbalanced datasets and immature evaluation practices currently challenge the applicability of universal machine-learned potentials to biomolecular simulations.
\end{abstract}

\begin{document}

\flushbottom
\maketitle

\thispagestyle{empty}

\section*{Introduction}

Molecular simulations offer atomic-level insights into the structure, dynamics, and interactions of biological systems.\cite{Karplus1990, Huggins2019} Their reliability depends on the accuracy of the underlying potential models, as even small errors can lead to unrealistic conformations.\cite{Chandrasekhar2003} Classical force fields (FFs), widely used in biomolecular simulations, are typically parameterized against first-principles calculations or experimental data. However, their simplified treatment of inter- and intramolecular interactions limits accuracy. Free energy differences between native and non-native protein states are often on the order of 5--15 kcal/mol.\cite{Makhatadze1995} Achieving this level of accuracy is essential in areas such as computational protein design, where the designed sequence must fold correctly and the native structure represents the global energy minimum.\cite{Bennett2023} Inaccurate FFs can, thus, lead to misfolded or non-functional conformations. Overall, the limited accuracy of classical FFs reduces their applicability in diverse areas, from modeling reactive processes to accurately predicting thermodynamic and kinetic properties.\cite{vanGunsteren2024}

Machine-learned (ML) potentials have emerged as a powerful alternative, offering accuracy comparable to that of first-principles approaches, such as density functional theory (DFT), at a fraction of the computational cost.\cite{Behler2007, Shapeev2016, Drautz2019, Zaverkin2020, Musaelian2023, Batzner2022, Batatia2022, Zaverkin2024b, Unke2021b, Anstine2025} These potentials have been successfully applied to a wide range of molecular and material systems.\cite{Friederich2021} However, their use in biological simulations remains limited, due to challenges in generating training sets with balanced and unbiased coverage across relevant compositional (atom types) and vibrational (atomic positions) degrees of freedom.\cite{Unke2024} Recent advances in universal ML potentials,\cite{Batatia2023, Yang2024} aimed at generalizing across relevant chemical and conformational spaces through training on large-scale datasets,\cite{Deng2023, Eastman2023b, Eastman2024, BarrosoLuque2024, Levine2025} have also promoted their development for biomolecular applications.\cite{Kovacs2025, Kabylda2025, Ple2025} However, generalization of these models to biomolecular systems under realistic conditions remains largely unexplored.

\begin{figure*}[htbp]
    \centering
	\includegraphics[width=0.795\linewidth]{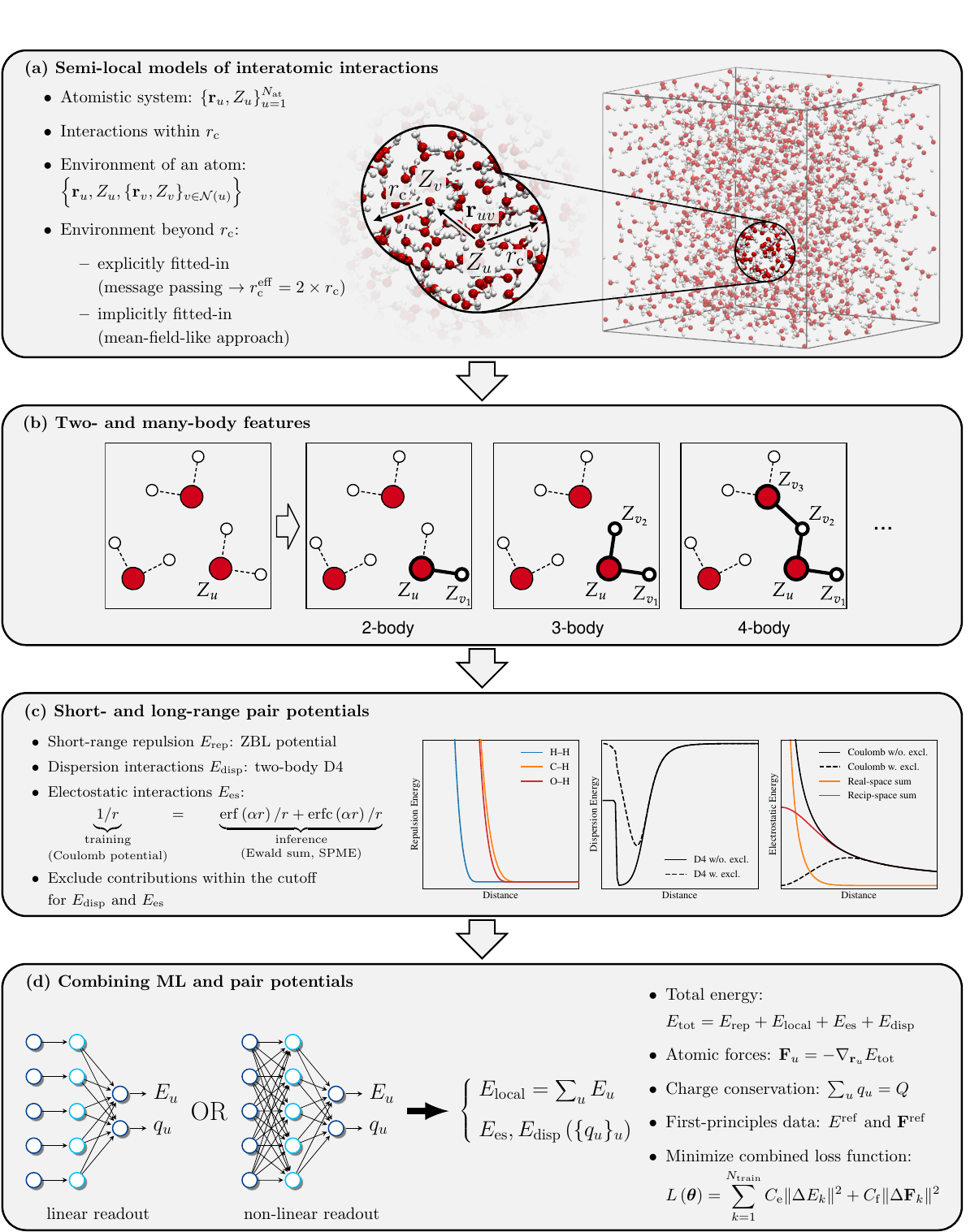}
	\caption{Schematic representation of ML potentials with and without explicit long-range dispersion and electrostatics. (a)~ML potentials often rely on (equivariant) message passing and explicitly capture interactions beyond the cutoff radius~$r_\mathrm{c}$. Interactions beyond the effective cutoff $r_\mathrm{c}^\mathrm{eff}$, with $r_\mathrm{c}^\mathrm{eff} = 2\times r_\mathrm{c}$ for two message-passing layers, are included implicitly. (b)~Atomic environments are encoded using many-body features such as distances (2-body), angles (3-body), and dihedrals (4-body), among others. These features depend on configurational and vibrational degrees of freedom. (c)~By incorporating analytic pair potentials, ML models more accurately capture short-range repulsion and long-range interactions with a characteristic power-law decay. For two-body dispersion, a short-range cutoff is typically sufficient due to fast decay. In contrast, electrostatic interactions often require treatment without a short-range cutoff or using Ewald summation or the SPME method, which involve real- and reciprocal-space cutoffs. Since ML potentials already model short-range interactions, contributions from long-range pair potentials within the cutoff radius $r_\mathrm{c}$ are often excluded. (d)~ML potentials employ linear and non-linear readout layers to derive atomic energies $E_u$ and partial charges $q_u$ from many-body features, training them with a combined loss function that includes reference energies $E^\mathrm{ref}$ and forces $\mathbf{F}^\mathrm{ref}$. Here, $\Delta E_k$ and $\Delta \mathbf{F}_k$ denote differences between predicted and reference total energies and forces for structure $k$ in the training set, and $\boldsymbol{\theta}$ represents the trainable parameters. To ensure transferability between the Coulomb potential and Ewald or SPME methods, ML models should be constrained to conserve total charge $Q$.}
	\label{fig:overview}
\end{figure*}

We aim to assess the applicability of ML potentials pre-trained on large-scale datasets to biomolecular simulations. We consider a representative class of ML potentials that construct geometric representations of atomic environments, capturing relevant compositional and vibrational degrees of freedom using (equivariant) message passing.\cite{Batzner2022, Batatia2022, Zaverkin2024b} As illustrated in \figref{fig:overview}~(a), these models learn atom-centered representations by iteratively processing local information, incorporating interactions beyond the cutoff radius. The expressive power of atom-centered representations depends on capturing symmetries of local environments and many-body correlations.\cite{Pozdnyakov2020, Joshi2023} A common approach embeds directional information from atomic environments into higher-dimensional tensor spaces and applies parameterized tensor products to compute many-body features,\cite{Shapeev2016, Drautz2019, Zaverkin2020, Musaelian2023, Batzner2022, Batatia2022, Zaverkin2024b} which are shown in \figref{fig:overview}~(b).

Long-range dispersion and electrostatics are fundamental to the structure, dynamics, and function of biological matter.\cite{Ren2012} Message-passing architectures extend the effective interaction range, yet we evaluate the impact of including explicit long-range interactions in biomolecular simulations. As illustrated in \figref{fig:overview}~(c), we combine ML potentials with the analytic two-body term of the D4 dispersion correction\cite{Caldeweyher2019} and the Coulomb potential, which rely on learnable partial charges. We further include an empirical repulsion potential to ensure correct short-range asymptotic behavior.

In contrast to previous work,\cite{Anstine2025, Unke2024, Kabylda2025} we do not impose a cutoff distance for electrostatics, either during training or simulations. We employ the smooth particle mesh Ewald (SPME) method\cite{Essmann1995} in bulk-system simulations, which enables us to exploit automatic differentiation capabilities of PyTorch.\cite{Christiansen2025} As shown in \figref{fig:overview}~(d), atomic energies and partial charges are derived from a shared message-passing representation using property-specific readout layers. The learnable partial charges are trained exclusively on total energies and atomic forces. However, unlike prior work,\cite{Cheng2025} we ensure transferability between the Coulomb potential used during training and the SPME method applied during simulations.

In this work, we use irreducible Cartesian tensor potentials (ICTPs),\cite{Zaverkin2024b} which extend the MACE architecture\cite{Batatia2022} to the Cartesian basis and represent the broad class of architectures discussed above. We compare ICTP models of varying sizes, including ICTP-LR(S), ICTP-LR(M), and ICTP-LR(L) with explicit long-range interactions, and short-range ICTP-SR models. These models are trained on SPICE-v2,\cite{Eastman2023b, Eastman2024} augmented with additional drug-like molecules and water clusters, both with and without \ce{Na+} and \ce{Cl-} ions. We benchmark their performance across in- and out-of-distribution test datasets, and in molecular simulations of biologically relevant systems.

We assess mean absolute errors (MAEs) and root mean square errors (RMSEs) in energies and forces on held-out test subsets of the augmented SPICE-v2 dataset (in-distribution), as well as on test-only datasets\cite{Eastman2024} and torsion profiles of drug-like molecules (out-of-distribution).\cite{Lahey2020, Rai2022} These datasets enable us to evaluate how model size, long-range electrostatics, and training data composition impact predictive accuracy, particularly in terms of generalization to larger molecular systems and modeling interactions in complex solvation environments.

We investigate whether results from test datasets extend to bulk systems by simulating pure liquid water and \ce{NaCl}-water mixtures, comparing predicted densities and radial distribution functions (RDFs) with experimental data. These simulations provide a computationally efficient yet sensitive benchmark for evaluating short-range accuracy, and the impact of long-range interactions and training data composition in modeling biologically relevant systems. We also study the alanine tripeptide (Ala3) in cationic and blocked forms, along with the mini-protein Trp-cage. These systems are small enough for extensive sampling, yet complex enough to assess the impact of model size and explicit electrostatics on conformational ensembles and free energy landscapes. We compute the vibrational power spectrum of Crambin to evaluate the models' accuracy on larger proteins. Finally, we compare ML potentials to classical FFs using errors relative to DFT calculations, highlighting tradeoffs between accuracy and computational efficiency.

\section*{Results}

\subsection*{Accuracy for the in- and out-of-distribution datasets}

\begin{figure*}[htbp]
	\includegraphics[width=\linewidth]{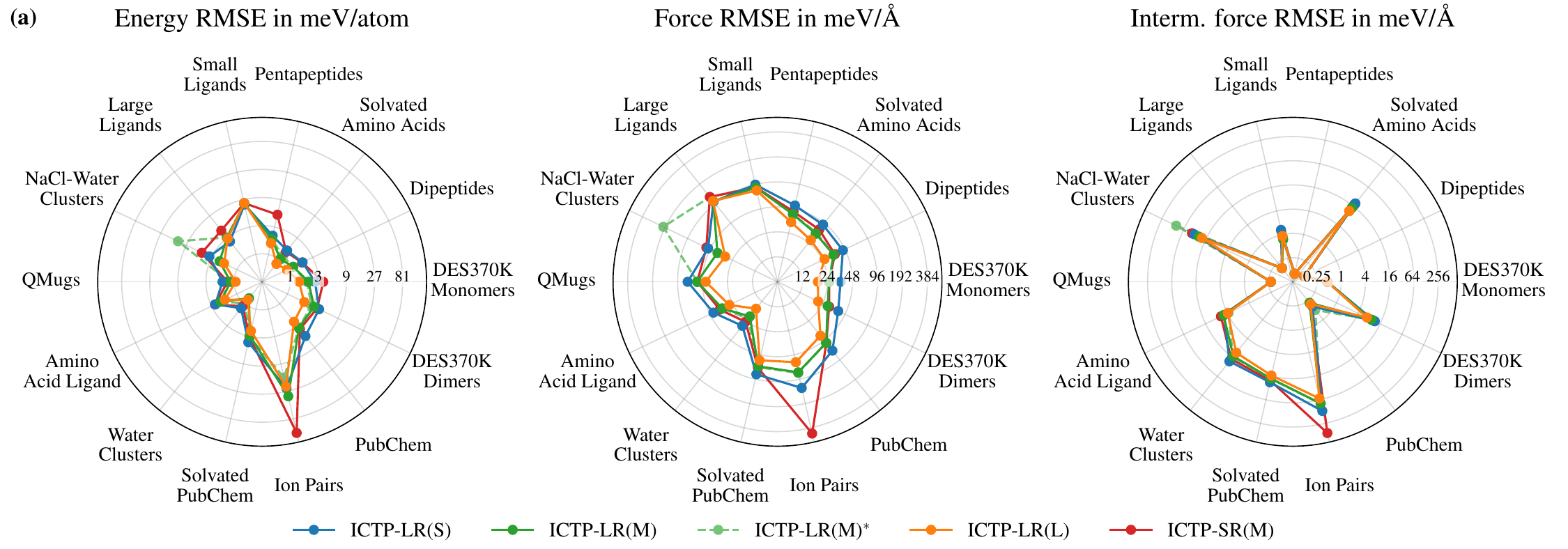}\vfill
    \includegraphics[width=\linewidth]{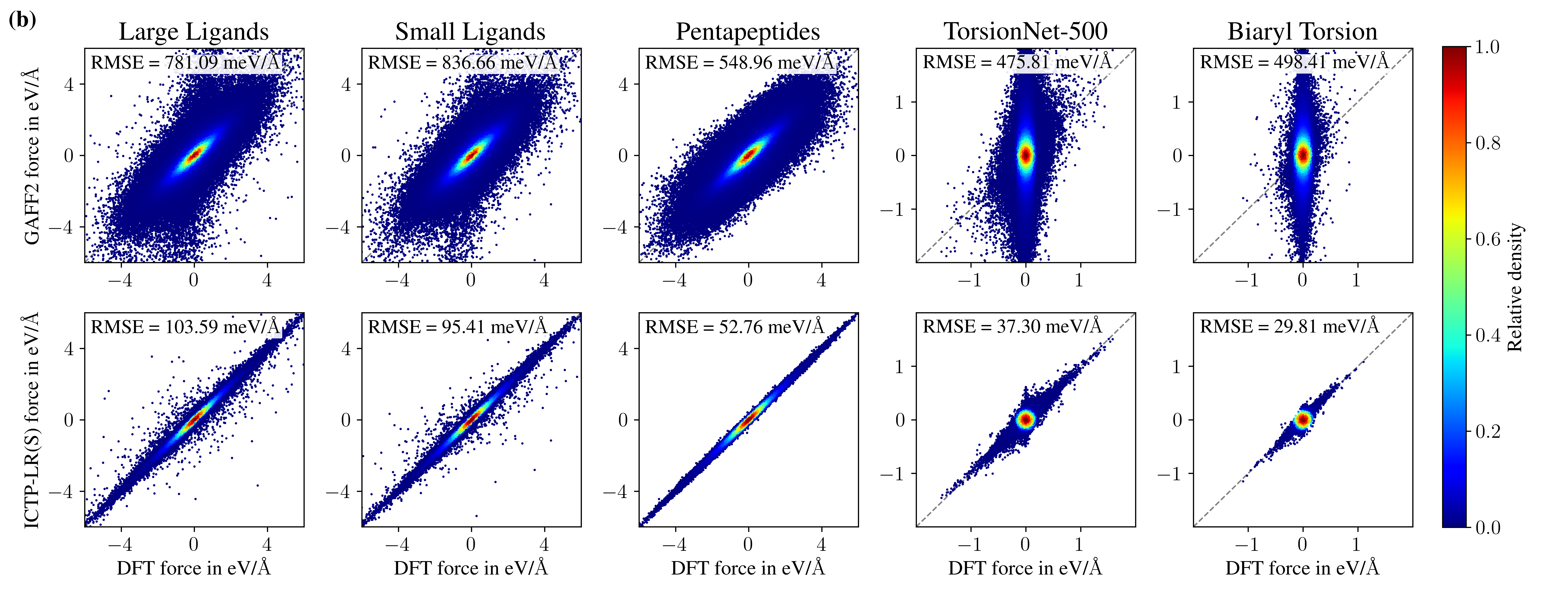}\vfill
    \includegraphics[width=\linewidth]{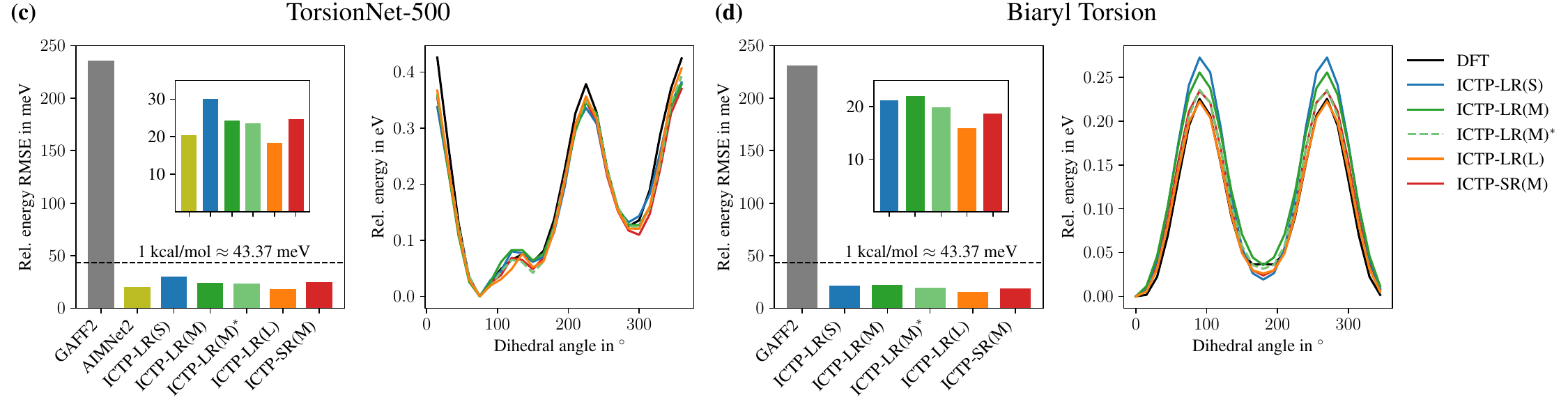}
	\caption{Performance of ICTP models across benchmark datasets. (a)~RMSEs in the predicted energies (left), forces (middle), and intermolecular forces (right) for the individual test datasets. The small values for intermolecular force errors observed in datasets composed of single molecules (PubChem, DES370K Monomers, Dipeptides, QMugs, Pentapeptides, Small Ligands, and Large Ligands) arise from DFT forces not summing to zero, whereas forces predicted by ICTP models do. For the Ion Pairs dataset, we provide force errors only for the $z$-component, as the $x$ and $y$ components are zero. The larger energy and force errors in the Large and Small Ligand datasets are primarily due to a few outliers; see (b) for details. (b)~Correlation between the predicted forces (by GAFF2 and ICTP-LR(S)) and DFT forces for the test-only datasets. The colors indicate the relative density as estimated by a kernel density estimator. (c, d; left)~RMSEs of relative energies for the TorsionNet-500 and Biaryl Torsion test datasets. (c, d; right)~Potential energy profiles for representative structures from the TorsionNet-500 and Biaryl Torsion test datasets. ICTP-LR(M)$^\ast$ corresponds to ICTP-LR(M) trained without the \ce{NaCl}-Water Clusters dataset.}
	\label{fig:test_errors}
\end{figure*}

\Figref{fig:test_errors}~(a) shows RMSEs for energies and forces computed on each held-out and test-only dataset. The test-only datasets include Pentapeptides, Small Ligands, and Large Ligands.\cite{Eastman2024} We also evaluate the intermolecular force errors,\cite{Kovacs2025} obtained by separating the force contributions to molecular translations and rotations. Overall, we observe that model accuracy consistently improves with increasing size, as demonstrated by the performance of ICTP-LR(S), ICTP-LR(M), and ICTP-LR(L).

ICTP-LR(L) achieves RMSEs on held-out datasets of 0.82--2.46~meV/atom for energy, 15.58--56.20~meV/\AA{} for forces, and 3.91--20.21~meV/\AA{} for intermolecular forces. For reference, the chemical accuracy limit is defined as 1~kcal/mol ($\approx$43.37~meV). These results exclude the Ion Pairs dataset, which contains only two atoms per structure and can yield higher per-atom energy errors. For this dataset, ICTP-LR(L) yields RMSEs of 22.21~meV/atom for energy and 59.29~meV/\AA{} for forces. In comparison, ICTP-LR(S) typically achieves RMSEs 1.5--2.0 times larger across datasets.

Comparing ICTP-LR(M) and ICTP-SR(M), we find similar performance on held-out datasets, except for Ion Pairs and \ce{NaCl}-Water Clusters, where ICTP-LR(M) performs better. This result suggests the importance of explicit long-range interactions in systems with unscreened charges and complex solvation environments. On test-only datasets, ICTP-LR(M) also yields lower energy RMSEs than ICTP-SR(M), indicating that incorporating explicit long-range interactions improves the generalization of ML potentials. In contrast, force RMSEs are less sensitive to explicit long-range interactions due to their faster decay and the 10~\AA{} effective cutoff of ICTP models.

ICTP-LR(M) and ICTP-LR(M)$^\ast$, trained with and without the \ce{NaCl}-Water Clusters dataset, respectively, yield similar RMSEs across held-out and test-only datasets. A notable difference appears only in the \ce{NaCl}-Water Clusters dataset itself. This result suggests that including \ce{NaCl}-water clusters is important for accurately modeling \ce{NaCl}-water mixtures and related systems.

To put our results into perspective, we provide force RMSEs for the test-only datasets obtained with GAFF2; see \figref{fig:test_errors}~(b). GAFF2 force errors are at least an order of magnitude higher than those from the ICTP models. We do not report energy errors, as only the relative energies are meaningful in this context and are discussed in the next section for torsion profiles of drug-like molecules.

\Figref{fig:test_errors}~(b) shows the force correlation between the GAFF2-predicted and DFT forces. GAFF2 forces are highly uncorrelated with DFT values and have a much broader distribution (larger RMSE) compared to the ICTP models. Our results suggest that classical FFs may have limited applicability for accurate property prediction. We acknowledge that parameter assignment in GAFF2 may introduce errors. However, we do not expect them to significantly impact our results.

MACE and Allegro,\cite{Kovacs2025, Tan2025} pre-trained on SPICE-v2, provide another meaningful baseline. The comparison with Allegro and MACE positions the ICTP models within the context of established approaches, rather than emphasizing their absolute accuracy. It supports a broader generalization of our findings in this and the following sections.

We compare the performance of Allegro and the ICTP models on the original, unrestricted test-only datasets.\cite{Eastman2024} For Allegro, force (energy) MAEs are 42--47~meV/\AA{} (220--2164~meV) for the smallest model and 27--36~meV/\AA{} (165--2231~meV) for the largest. For the ICTP models, force RMSEs are generally on par with Allegro, typically within a factor of 1.2, while energy RMSEs are 2 to 20 times lower. These results hold for models with and without explicit long-range electrostatics. For the held-out datasets, ICTP-LR(M) achieves MAEs comparable to those reported for MACE-OFF24(M), typically within a factor of 2.5 and below the chemical accuracy limit. Further details are provided in Supplementary Tables 3 and 4.

\subsection*{Torsion profiles of drug-like molecules}

Figures~\ref{fig:test_errors}~(c,d) evaluate ICTP models using torsion profiles of drug-like molecules from the TorsionNet-500\cite{Rai2022} and Biaryl Torsion datasets.\cite{Lahey2020} TorsionNet-500 includes profiles for 500 molecules spanning a broad range of pharmaceutically relevant chemical space. Biaryl Torsion includes torsional energy profiles for about 100 biaryl fragments, in which a rotatable bond connects two aromatic rings. Following Ref.~\citenum{Kovacs2025}, we use energies and forces recomputed using the DFT settings of SPICE-v2.

Figures~\ref{fig:test_errors}~(c, d; left) show the relative energy RMSEs of the torsion profiles for the ICTP models differing by the inclusion of explicit long-range interactions and the \ce{NaCl}-Water Cluster dataset. We also include GAFF2 and AIMNet2\cite{Anstine2025} for comparison. We again observe systematic improvements in accuracy as the size of the ICTP models increases. The inclusion of long-range interactions has no noticeable impact on accuracy, due to the relatively small size of the molecules and the sufficiently large receptive field of the ICTP models. All ICTP models, regardless of size, achieve errors below the chemical accuracy limit and outperform GAFF2 by at least an order of magnitude. Figures~\ref{fig:test_errors}~(c, d; right) demonstrate the ability of the ICTP models to capture complex potential energy profiles, including regions far from equilibrium geometries.

\subsection*{Bulk water with various salt concentrations}

\begin{figure*}[t!]
	\includegraphics[width=0.3333333\linewidth]{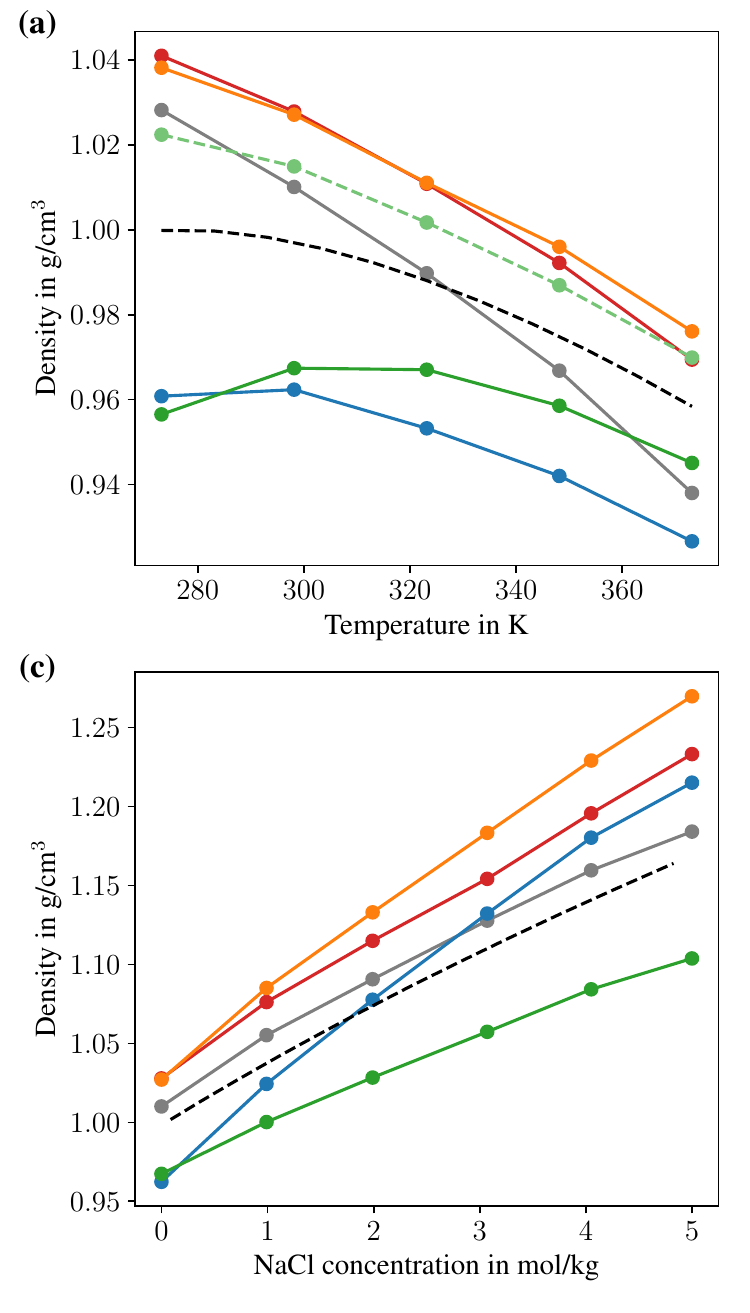}\hfill
    \includegraphics[width=0.6666666\linewidth]{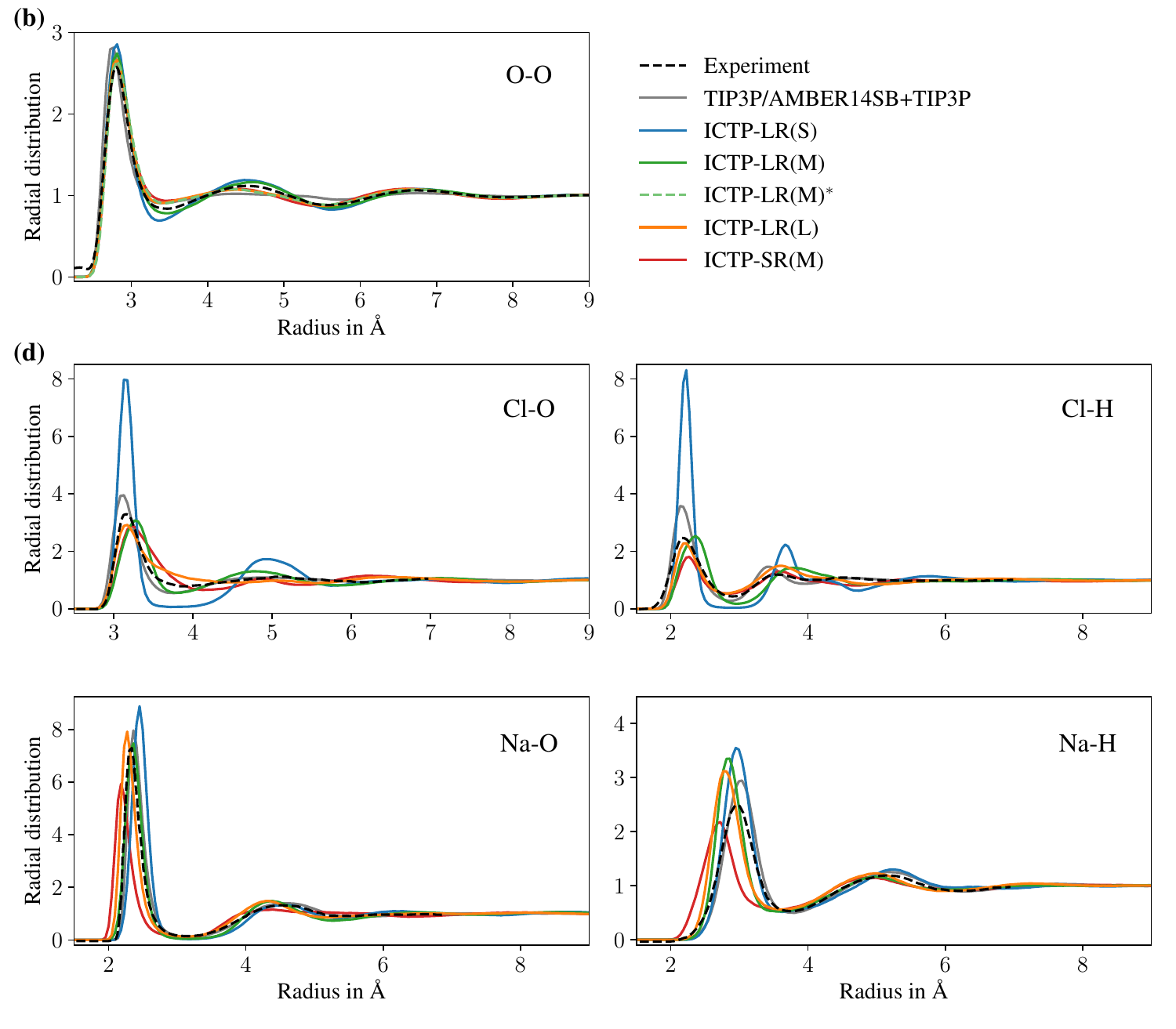}
	\caption{Properties of pure liquid water and \ce{NaCl}–water mixtures. (a)~Density of pure liquid water as a function of temperature. (b)~The O--O RDF of pure liquid water at 298.15~K. (c)~Density of the \ce{NaCl}–water mixture as a function of \ce{NaCl} concentration. (d)~RDFs of \ce{Na^+} and \ce{Cl^-} ions in the \ce{NaCl}–water mixture at 0.99~mol/kg and 298.15~K. Experimental densities are from Ref.~\citenum{Lide2005}, the O--O RDF is from Ref.~\citenum{Skinner2013}, and the Cl--O, Cl--H, Na--O, and Na--H RDFs at 0.67~mol/kg are from Ref.~\citenum{Mancinelli2007}. ICTP-LR(M)$^\ast$ corresponds to ICTP-LR(M) trained without the \ce{NaCl}–Water Clusters dataset.}
	\label{fig:water-nacl_solution-results}
\end{figure*}

Figures~\ref{fig:water-nacl_solution-results}~(a,c) show the density of pure liquid water as a function of temperature at 1~bar, and of \ce{NaCl}-water mixtures as a function of \ce{NaCl} concentration at 298.15~K and 1~bar. Densities are computed using the TIP3P and ICTP models and compared with corresponding experimental data. In the isothermal-isobaric ($NpT$) ensemble, density becomes an observable sensitive to small errors in intermolecular interactions,\cite{Magdau2023} making it an excellent metric for evaluating the accuracy of ML potentials.

At 298.15~K, the predicted water density from ICTP-LR(S) is within 3.5~\% of the experimental value. ICTP-LR(M) and ICTP-LR(L) perform slightly better, with deviations within 3.0~\%. These results suggest a systematic improvement with increasing model size. However, ICTP-LR(L) performs slightly worse on average than ICTP-LR(M), predicting densities within 2.6~\% of the actual values, averaged over the temperature range, compared to 2.5~\% for ICTP-LR(M). Overall, when examining densities at individual temperatures, we find no consistent improvement with increasing model size. Similar behavior is observed for \ce{NaCl}-water mixtures, with ICTP-LR(S) outperforming the larger models.

TIP3P predicts the density of water within 1.3~\% of the experimental value at 298.15~K and outperforms the ICTP models. Unlike prior work,\cite{Vega2011} we found that the flexible TIP3P model, used to ensure consistency with the ICTP models, overestimates the density at 298.15~K rather than underestimating it. AMBER14SB+TIP3P yields \ce{NaCl}-water mixture densities that are on average comparable to ICTP-LR(S) and better than those from the larger models. These results demonstrate the benefit of parameterizing models to reproduce experimental properties, even if such models correlate only weakly with DFT forces. Recent work shows that incorporating experimental data into ML models also improves their accuracy in MD simulations.\cite{Matin2024}

Explicit long-range interactions do not improve the accuracy of predicted water densities compared to short-range models. This result agrees with previous work, which reports no significant performance differences in water models with or without explicit electrostatics when using a cutoff of around 6.35~\AA.\cite{Morawietz2016} ICTP-SR(M), with an effective cutoff of 10~\AA{}, yields densities within 2.5~\% averaged over the temperature range. Furthermore, ICTP-LR(M) tends to underestimate density, while ICTP-SR(M) overestimates it, resulting in differences of up to 8~\%. Similar behavior is observed for \ce{NaCl}-water mixtures.

The inclusion of the \ce{NaCl}-Water Clusters dataset significantly affects model performance. ICTP-LR(M)$^\ast$, trained without it, outperforms ICTP-LR(M) and predicts water densities within 1.6~\% of the experimental values averaged over the temperature range. Similar to ICTP-SR(M), ICTP-LR(M)$^\ast$ overestimates water densities. Despite its improved accuracy for pure water, ICTP-LR(M)$^\ast$ fails in longer MD simulations of \ce{NaCl}-water mixtures. This sensitivity to a small subset raises the question of whether it is a consequence of imbalanced data generation strategies used in large-scale datasets.

For comparison, MACE-OFF24(M), with an effective cutoff of 12~\AA, predicts water density within 2~\% of the experimental value at 298.15~K.\cite{Kovacs2025} MACE-OFF23(M), with a 10~\AA{} cutoff, overestimates it by about 20~\%, which contrasts with our results and prior work.\cite{Morawietz2016} We do not compare our results with the recent FeNNix-Bio1 models,\cite{Ple2025} as their reported densities are computed using non-classical MD that incorporates nuclear quantum effects.

Figures~\ref{fig:water-nacl_solution-results}~(b,d) show the O--O RDF for pure liquid water, as well as the Cl--O, Cl--H, Na--O, and Na--H RDFs for the \ce{NaCl}-water mixture at 0.99~mol/kg. All ICTP models perform well on the O--O RDF, with the ICTP-LR(S) model providing slightly overstructured water. Differences in O--O RDFs across models correlate with the density trends at 298.15~K. While ICTP-LR(S) predicts more accurate densities for \ce{NaCl}-water mixtures than the larger models, it shows stronger overstructuring in the Cl--O and Cl--H RDFs. Overall, the RDFs of the \ce{NaCl}-water mixture at 0.99~mol/kg show a more systematic improvement with increasing model sizes, in contrast to the density trends in \figref{fig:water-nacl_solution-results}~(c).

\subsection*{Alanine tripeptide in aqueous solutions}

\begin{figure*}[htbp]
	\includegraphics[width=\linewidth]{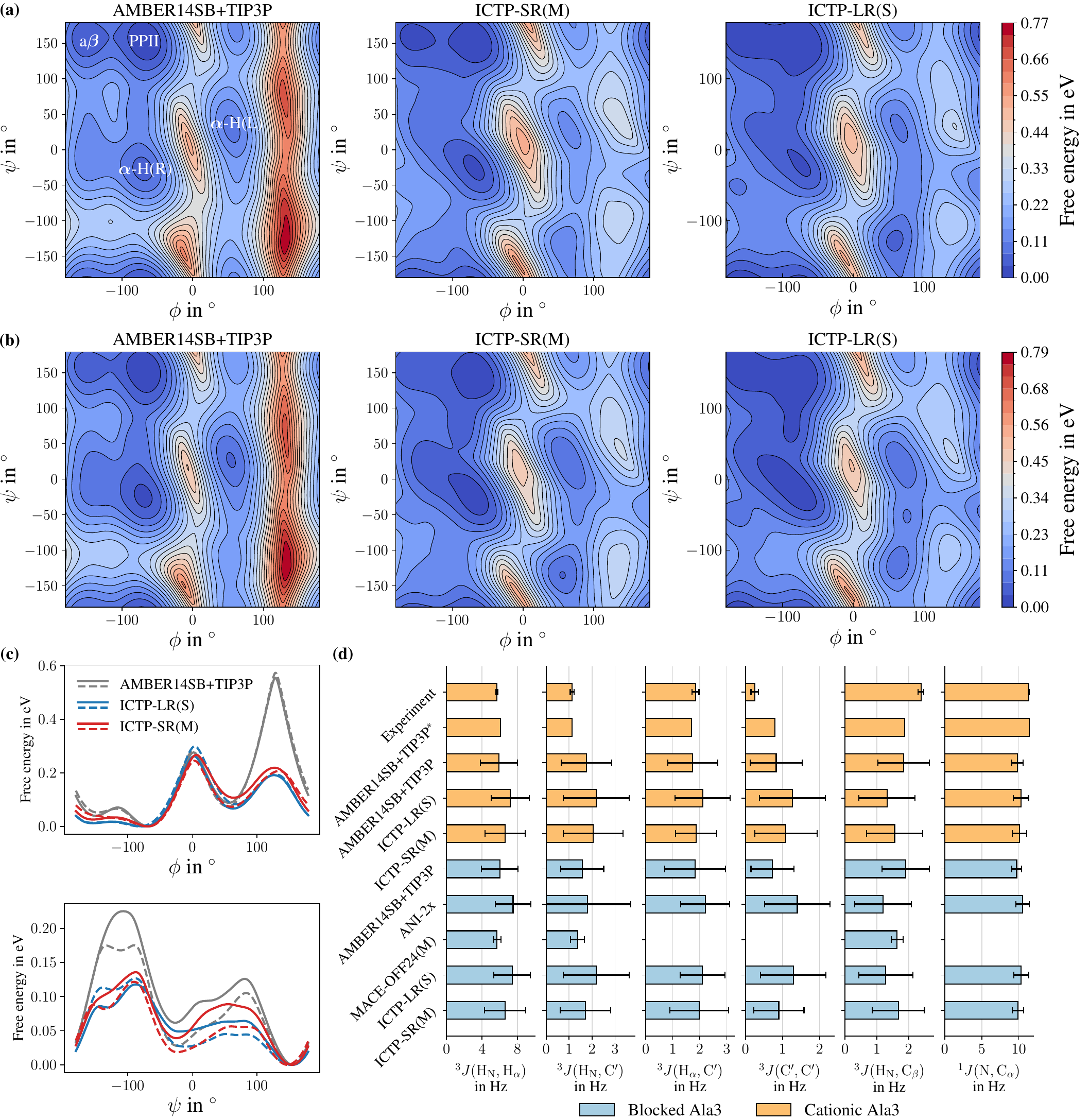}
	\caption{FES and $J$-coupling constants of Ala3 in its cationic form (in explicit \ce{NaCl}-water mixture) and in its blocked form with neutral \ce{N}- and \ce{C}-termini (in explicit water). (a, b) Two-dimensional FES as a function of the backbone dihedral angles $\phi$ and $\psi$ at 298.15~K and 1~bar, computed with AMBER14SB+TIP3P, ICTP-SR(M), and ICTP-LR(S) for the cationic (a) and blocked (b) forms. (c) One-dimensional FES for $\phi$ and $\psi$. Solid lines represent the cationic form, while dashed lines correspond to the blocked form. (d) $J$-coupling constants for cationic and blocked Ala3 derived from dihedral angle distributions. For ICTP-LR(S), ICTP-SR(M), and AMBER14SB+TIP3P, expectation values and standard deviations are computed by integrating over reweighted dihedral angle probability densities obtained from metadynamics simulations. For ANI-2x, the corresponding values are computed from conformations sampled during 10~ns of unbiased $NVT$ dynamics. In both cases, the standard deviation reflects the variability of the $J$-coupling constants arising from fluctuations in backbone dihedral angles. For MACE-OFF24(M), the $J$-coupling constants are obtained from 20~ns of unbiased $NpT$ simulations, and standard errors of the mean are obtained from block averaging. Experimental, AMBER14SB+TIP3P$^\ast$, and ANI-2x values are taken from Refs.~\citenum{Graf2007, Zhang2020, Rosenberger2021}.}
	\label{fig:ala3-results}
\end{figure*}

Figures~\ref{fig:ala3-results}~(a)--(c) present the FES of cationic and blocked Ala3 in explicit aqueous solution, computed using AMBER14SB+TIP3P, ICTP-SR(M), and ICTP-LR(S). While ICTP-SR(M) implicitly accounts for interactions beyond a 10~\AA{} cutoff, ICTP-LR(S) explicitly incorporates long-range dispersion and electrostatics. This choice of ML potentials and peptide forms allows us to investigate the impact of explicit long-range interactions and model capacity on the predicted conformational landscapes.

All models identify four local minima: the antiparallel $\beta$-sheet ($\phi < -120^\circ, \psi > 120^\circ$), a right-handed $\alpha$-helix ($\phi = -60^\circ, \psi > -60^\circ$), the corresponding left-handed $\alpha$-helix ($\phi = 60^\circ, \psi < 60^\circ$), and a polyproline II (PPII)-type structure ($\phi = -60^\circ, \psi > 120^\circ$). Their relative depths agree between ICTP-SR(M) and AMBER14SB+TIP3P within 15 meV for blocked and 35 meV for cationic Ala3. For the cationic form, ICTP-SR(M) predicts the right-handed $\alpha$-helix 34~meV lower and the left-handed $\alpha$-helix 35~meV higher in energy than AMBER14SB+TIP3P. We also find differences in barrier heights, consistent with previous work.\cite{Kovacs2025}

ICTP-LR(S) generally yields similar results to ICTP-SR(M) but predicts the antiparallel $\beta$-sheet to be nearly degenerate with the PPII-type structure (within 5~meV). This result contradicts the well-established intrinsic preference of unfolded peptides for the PPII conformation, which is largely independent of the protonation state.\cite{Toal2013} The conformational preferences predicted by AMBER14SB+TIP3P and the ICTP models remain essentially unchanged between the two forms of Ala3, regardless of whether explicit long-range interactions are included.

\Figref{fig:ala3-results}~(d) shows differences in conformational distributions between AMBER14SB+TIP3P and the ICTP models using $J$-coupling constants. These constants also enable direct comparison with experimental data and ML potentials such as ANI-2x and MACE-OFF24(M). Following prior work,\cite{Zhang2020} we report five $^3J$ and one $^1J$ constants dependent on the central $\phi$ dihedral angle.

We compute $J$-coupling constants by integrating over reweighted dihedral angle probability densities obtained from 60~ns metadynamics simulations. While the total simulation time of 60~ns may impose some limitations, it is sufficient for meaningful comparisons between models. For example, for cationic Ala3, the values obtained with AMBER14SB+TIP3P agree with prior work,\cite{Zhang2020} with minor differences attributed to the flexible water model used in this work.

ICTP-SR(M) more closely reproduces experimental $J$-coupling data than its smaller counterpart ICTP-LR(S). We argue that the primary differences between ICTP-LR(S) and ICTP-SR(M) stem from their differing accuracy in describing short-range interactions, which directly influences how each model represents the torsional free energy surface. AMBER14SB+TIP3P often outperforms the ICTP models, again highlighting the benefits of training parameterized models to match experimental observables.

MACE-OFF24(M) constants for blocked Ala3 align more closely with experimental data for cationic Ala3 than those from the ICTP models. However, these differences may arise from how $J$-coupling constants are computed. To assess this dependence, we performed a 20~ns unbiased $NpT$ simulation with ICTP-LR(S) and computed $J$-coupling constants using block averaging. This procedure yielded $^3J(\mathrm{H}_\mathrm{N},\mathrm{H}_\alpha) = 6.79 \pm 0.24$~Hz compared to $7.40 \pm 2.06$~Hz obtained from metadynamics-based distributions, with other constants differing by 0.1--0.35~Hz. Since we consider the metadynamics-based approach more robust, no additional unbiased simulations were performed.

\subsection*{Small proteins---Trp-cage and Crambin}

\begin{figure*}[t!]
    \centering
	\includegraphics[width=\linewidth]{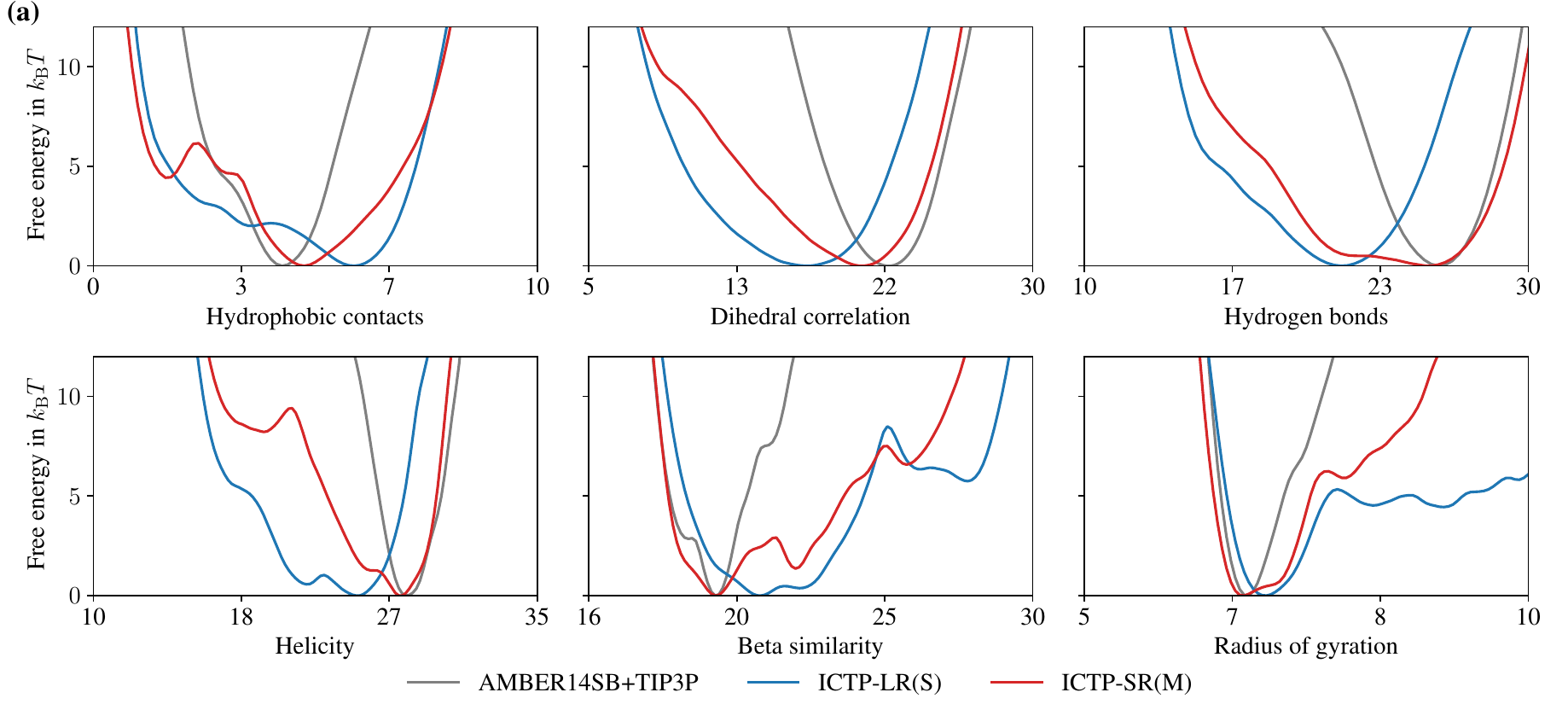}\vfill
    \includegraphics[width=\linewidth]{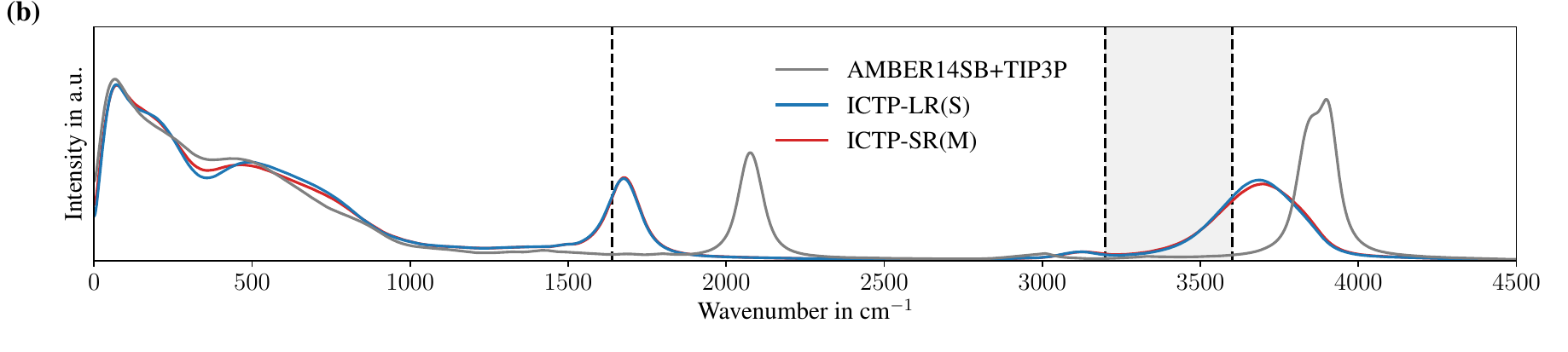}
	\caption{Simulation results for the mini-protein Trp-cage and Crambin in an explicit \ce{NaCl}-water mixture using AMBER14SB+TIP3P, ICTP-SR(M), and ICTP-LR(S). (a) One-dimensional FES of Trp-cage at 298.15~K and 1~bar for the six monitored CVs: the number of C$_\gamma$-hydrophobic contacts, the dihedral correlation, the number of backbone hydrogen bonds, the $\alpha$- and $\beta$-dihedral fractions, and the C$_\alpha$-radius of gyration. (b) Vibrational power spectrum of Crambin obtained from the last 125~ps of 1.2~ns MD, recorded with a time resolution of 0.5~fs. Black lines denote experimental peaks at 1640 cm$^{-1}$ and the range 3200--3600 cm$^{-1}$, corresponding to bending and stretching vibrations of water molecules. The grey area additionally highlights the stretching region.}
	\label{fig:trp-crambin-results}
\end{figure*}

\Figref{fig:trp-crambin-results}~(a) shows the FES of the mini-protein Trp-cage along six collective variables (CVs), and compares ICTP-SR(M), ICTP-LR(S), and AMBER14SB+TIP3P. To reduce computational cost, we used an integration time step of 1~fs instead of the 0.5~fs used in previous sections. A total of 60~ns of metadynamics was performed, which may limit quantitative FES estimations. Therefore, this section focuses on qualitative comparisons to assess the impact of architectural choices on conformational sampling.

The FES reveal differences in the conformational ensembles of Trp-cage predicted by each model. ICTP-SR(M) agrees more closely with AMBER14SB+TIP3P than ICTP-LR(S) in the positions of the free energy minima across all six CVs. The larger shifts predicted by ICTP-LR(S) may arise from the inclusion of explicit long-range interactions and less accurate modeling of short-range interactions compared to ICTP-SR(M). However, the origin of these shifts cannot be unambiguously identified without reference DFT calculations, which are essentially inaccessible, or an accurate, system-specific ML potential.

The FES shape is more consistent across the ICTP models, which exhibit greater variance along the CVs compared to AMBER14SB+TIP3P. The broader basins observed with the ICTP models suggest increased conformational flexibility. This trend aligns with previous findings from time-lagged root mean squared deviation analyses of Crambin.\cite{Unke2024, Kabylda2025} The restricted conformational variability of biomolecules with classical FFs can be attributed to the limited treatment of protein-solvent dispersion and polarization effects in commonly used water models.\cite{Piana2015}

The most notable difference between the ICTP models, aside from the shift in free energy minima, appears in the FES of the radius of gyration. ICTP-LR(S) spans a broader range of values and reveals intermediate states not present in ICTP-SR(M) or AMBER14SB+TIP3P. Overall, the FES predicted by each model is consistent with experimental\cite{Qiu2002, Neuweiler2005} and computational studies,\cite{Zhou2003, Kannan2014} supporting a two-state folding behavior dominated by the folded state but not precluding low-population intermediates along the folding pathway.\cite{Brockwell2007} However, we cannot unambiguously attribute the intermediates predicted by ICTP-LR(S) to the inclusion of explicit long-range interactions. Still, our findings suggest that explicit long-range electrostatics may be important for capturing the full conformational variability of Trp-cage.

In addition to the increased conformational flexibility of Trp-cage with the ICTP models, we observed proton transfer events between protonated nitrogen groups (\ce{-NH3^+} and \ce{=NH2^+}) and nearby deprotonated carboxylates (\ce{-COO^-}). Since the involved atoms were not used in the definition of the CVs, we do not expect these reactions to significantly impact the obtained FES. While a quantitative analysis of proton transfer events is beyond the scope of this work, their occurrence demonstrates the potential of ML potentials to reveal insights not accessible with classical FFs.

\Figref{fig:trp-crambin-results}~(b) shows the vibrational power spectrum of the larger protein Crambin, which agrees well with previous studies.\cite{Unke2024, Kovacs2025, Kabylda2025} The spectra predicted by the ICTP models are in close agreement, with no visible differences arising from the treatment of short- or long-range interactions. The characteristic water peaks at 1640~cm$^{-1}$ and 3200--3600~cm$^{-1}$ are well reproduced by the ICTP models. In contrast, the corresponding peaks predicted by AMBER14SB+TIP3P are blue-shifted and narrower. The broader peaks in the ICTP spectra suggest that intermolecular interactions have a stronger influence on the corresponding frequencies than observed with AMBER14SB+TIP3P.

We identified two distinct peaks in the low-frequency region of the spectrum. The dominant peak at about 60~cm$^{-1}$ corresponds to localized internal side-chain fluctuations, while the peak at about 220~cm$^{-1}$ is attributed to water intermolecular vibrational modes.\cite{Woods2014} Fully converging the THz region of the spectrum, which would enable quantitative comparison with the experiment, would require significantly longer simulations. Similar to prior work,\cite{Kovacs2025} the spectrum from the final 1.0~ns of a 1.2~ns long MD was not significantly different from that of the final 125~ps.

\subsection*{Inference times}

\begin{figure}[t!]
    \centering
    \includegraphics[width=\linewidth]{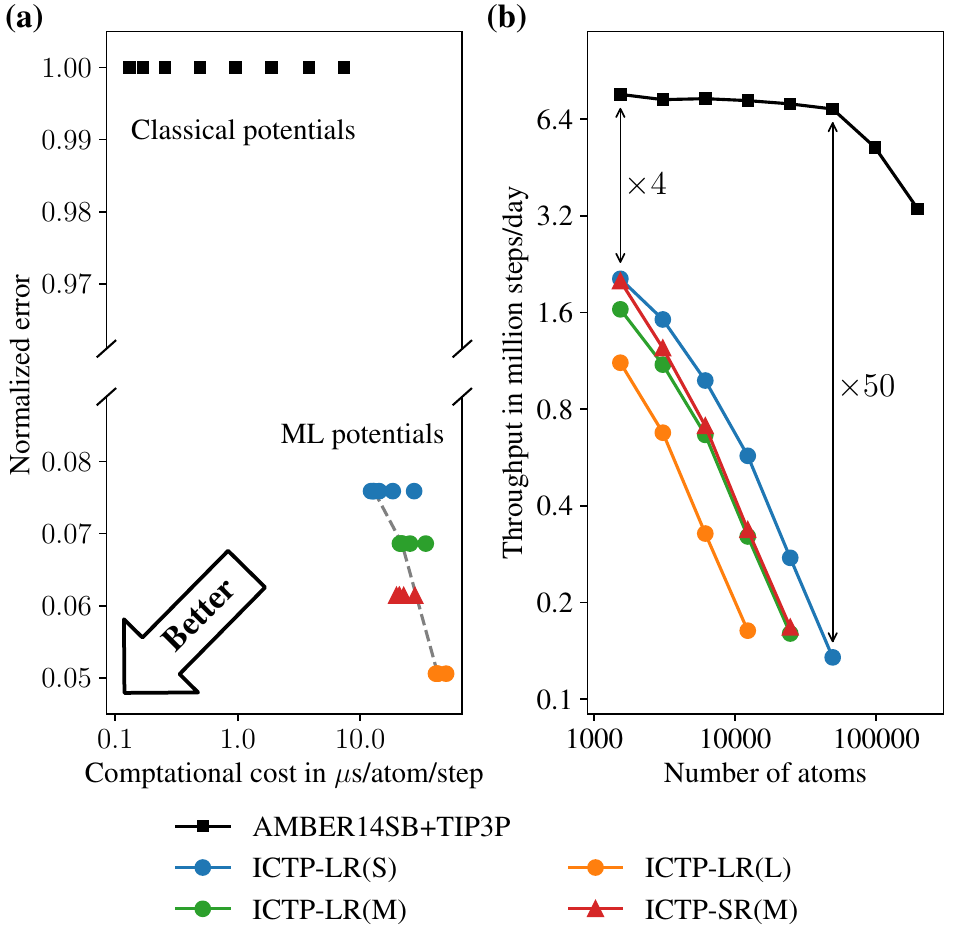}
	\caption{Inference time and throughput performance evaluated for the ICTP models and AMBER14SB+TIP3P. (a) The tradeoff between computational cost and normalized error (\eqref{eq:norm_error}). For the normalized error, the relative energy and force RMSEs obtained for the Biaryl Torsion dataset were used. For each model, multiple data points correspond to different system sizes, demonstrating the dependence of inference time on system size. (b) Throughput performance for various system sizes.}
	\label{fig:inference-results}
\end{figure}

\Figref{fig:inference-results}~(a) shows the inference time of the ICTP models in comparison to classical FFs (implemented in PyTorch) and in relation to the achievable accuracy with respect to DFT. \Figref{fig:inference-results}~(b) presents throughput performance as a function of system size. All results are from MD simulations performed with DIMOS\cite{Christiansen2025} in the canonical ($NVT$) ensemble at 298.15 K, using an integration time step of 0.5~fs over 1000 steps.

Inference time increases with model size, while accuracy improves at a slower rate. This accuracy-efficiency tradeoff favors smaller models, highlighting the need to improve their ability to learn more effectively from large-scale datasets. Including explicit long-range electrostatics through SPME has minimal impact on computational cost. The overhead is noticeable in small systems but negligible in larger ones, supporting the routine use of explicit long-range electrostatics with ML potentials.

ML potentials are slower than classical FFs, with the small ICTP-LR(S) model being 4 to 50 times slower than AMBER14SB+TIP3P, depending on system size. However, the speed of classical potentials is accompanied by higher errors compared to reference DFT, typically at least an order of magnitude larger in the normalized error compared to ML potentials.

\section*{Discussion}

Advances in universal ML potentials have been driven by the availability of large-scale datasets and the development of increasingly expressive model architectures. The accuracy of these potentials in biomolecular applications is typically assessed using energy and force RMSEs on benchmark datasets, and in molecular simulations relying on comparison with experimental data. In the absence of DFT-level simulation data or other high-quality baselines, a more reliable evaluation of accuracy should consider the dependence of RMSEs and simulation results on model expressivity, defined in this work by model size and the inclusion of explicit long-range interactions. In this context, we present the first systematic exploration of the applicability limits of universal ML potentials in biomolecular simulations, assessing the impact of model size, training data composition, and electrostatic treatment.

Our results show that RMSEs on benchmark datasets systematically improve with increasing model size. Incorporating explicit long-range interactions, even in ML potentials with an effective cutoff radius of 10~\AA, further enhances the model's generalization capability. These improvements, however, do not translate into systematic changes in predicted physical observables. In simulations of pure liquid water and the \ce{NaCl}-water mixture, larger models do not consistently outperform smaller ones when compared to the experimental data, and increasing model size does not lead to systematic convergence in predictions. For example, smaller models yield more accurate densities, while larger models more reliably reproduce experimental RDFs.

Including explicit long-range electrostatics also did not improve the accuracy of predicted densities and RDFs of pure liquid water and the \ce{NaCl}-water mixture. For pure liquid water, these interactions are expected to be less relevant, particularly in ML potentials with an effective cutoff larger than 6.35~\AA.\cite{Morawietz2016} However, we found that the model with electrostatics underestimates the density, whereas the model without electrostatics overestimates it. This behavior cannot be systematically assessed by comparison with experimental data alone or across different models.

Similar results were obtained for Ala3 and Crambin, with no evidence that explicit long-range electrostatics improve the accuracy of predicted properties. ICTP-SR(M) showed better agreement with experimental $J$-coupling constants than ICTP-LR(S) for Ala3, while no relevant differences were observed between models for Crambin. For Trp-cage, ML potentials systematically yielded greater conformational variability than classical FFs, with ICTP-LR(S) revealing low-population intermediates in the FES of the radius of gyration. This observation was identified as a potential advantage of explicitly modeling long-range electrostatics, but the lack of DFT-level benchmarks makes it difficult to quantify.

The generation of large-scale datasets generally lacks strategies to ensure balanced and unbiased coverage across relevant compositional and vibrational spaces. Such coverage is essential to achieve uniform accuracy of ML potentials and to prevent degradation of predictive accuracy in specific regions as the dataset grows. Our results demonstrate, for example, that predicted densities for pure liquid water depend on the composition of the training data. We expect such biases to persist even after fine-tuning universal ML potentials on smaller, specialized datasets. Advanced active learning offers a promising strategy to enhance coverage by allowing datasets to grow adaptively with the learning task.\cite{Zaverkin2024a} Still, this approach does not guarantee the construction of truly universal datasets, as their composition may remain biased by the specific model used to guide data generation and selection.\cite{Niblett2025}

In summary, our results suggest that immature evaluation practices and imbalanced data generation strategies currently challenge the reliability and applicability of universal ML potentials in realistic biomolecular settings. Among these challenges, the unfavorable tradeoff between efficiency and accuracy in large ML potentials appears the most tractable. Improving accuracy of smaller models through approaches such as knowledge distillation offers a promising direction.\cite{Gardner2025}

\section*{Methods}

\subsection*{Local machine-learned potentials}

A structure is defined as $S = \{\mathbf{r}_u, Z_u\}_{u=1}^{N_\mathrm{at}}$, where $\mathbf{r}_u \in \mathbb{R}^3$ and $Z_u \in \mathbb{N}$ denote the coordinates and atomic numbers of $N_\mathrm{at}$ atoms. The structure $S$ is mapped to a scalar energy $E$ through a function $f_{\boldsymbol{\theta}}: S \mapsto E \in \mathbb{R}$, parameterized by $\boldsymbol{\theta}$. Assuming locality of interatomic interactions, we split the total energy into a sum over individual atomic contributions\cite{Behler2007}
\begin{equation}
    \label{eq:energy-model}
    E\left(S, \boldsymbol{\theta}\right) = \sum_{u=1}^{N_\mathrm{at}} E_u\left(S_u, \boldsymbol{\theta}\right),
\end{equation}
where $S_u$ is the local environment of atom $u$ within a cutoff radius $r_\mathrm{c}$. Forces are obtained as $\mathbf{F}_u = -\nabla_{\mathbf{r}_u} E$. The parameters $\boldsymbol{\theta}$ are learned from datasets containing energies and forces.

We focus on message-passing architectures that represent structures as graphs in a three-dimensional Euclidean space. Atoms are treated as nodes, with an edge $\{u, v\}$ connecting two atoms $u$ and $v$ if $\lVert \mathbf{r}_u - \mathbf{r}_v \rVert_2 \leq r_\mathrm{c}$. Atom-centered representations are learned through iterative processing of local information, capturing many-body correlations and interactions beyond the cutoff $r_\mathrm{c}$.

\subsection*{Irreducible Cartesian tensor potential---ICTP}

ICTP generalizes concepts from earlier Cartesian tensor-based models\cite{Zaverkin2020} and extends the MACE architecture\cite{Batatia2022} to the Cartesian basis.\cite{Zaverkin2024b} It constructs atom-centered representations invariant to global rotations and translations, and permutations of atoms of the same species, using a rotation-equivariant message-passing based on irreducible Cartesian tensors. These tensors are used to represent node features and embed the directional information from normalized distance vectors $\hat{\mathbf{r}}_{uv}$ within a cutoff radius $r_\mathrm{c}$ (see \figref{fig:overview}~(a)), with $\hat{\mathbf{r}}_{uv} = \mathbf{r}_{uv} / \lVert\mathbf{r}_{uv}\rVert_2$. In this work, we use a cutoff of $r_\mathrm{c} = 5~\AA$.

In each message-passing layer, two-body features are constructed as tensor products between directional embeddings and node features, parameterized by learnable radial functions dependent on distances $\lVert\mathbf{r}_{uv}\rVert_2$. Node features in the first layer are initialized using invariant embeddings of the atomic species $Z_u$, augmented by invariant embeddings of the total charge $Q$ computed using an attention-like mechanism.\cite{Unke2021b}

Higher-body-order correlations are captured by successive tensor products of the two-body features, yielding many-body features (see \figref{fig:overview}~(b)) without requiring explicit summation over atom triplets, quadruplets, or higher-order tuples. These features are linearly combined using species- and tensor-rank-specific weights and then passed through an update step with a residual connection, yielding the node features of the next layer. After each message-passing layer, a readout layer is applied to the invariant components of the node features (see \figref{fig:overview}~(d)). The atomic energies $E_u\left(S_u,\boldsymbol{\theta}\right)$ and partial charges $q_u\left(S_u,\boldsymbol{\theta}\right)$ are then obtained by summing over the outputs from all message-passing layers.

Increasing the rank of directional embeddings and node features, or the correlation order, can significantly raise the computational cost while offering only marginal accuracy gains. However, tensors with a maximal rank of two are generally sufficient to represent node features and embed directional information, as local symmetries of atomic environments are typically lifted in atomistic simulations. Higher-body-order correlations per message-passing layer can also be limited to three-body interactions, computed using a single tensor product between two-body features. With two message-passing layers, the effective body order increases to seven,\cite{Zaverkin2024b} which is sufficient to resolve degeneracies in common local environments. In this work, model capacity is varied exclusively by changing the number of feature channels. We use 64, 128, or 256 channels for ICTP-LR(S), ICTP-LR(M), and ICTP-LR(L) or their short-range counterparts.

All parameters of the ICTP models are optimized by minimizing the combined squared loss on training data $D_\mathrm{train} = \left(X_\mathrm{train}, Y_\mathrm{train}\right)$, where $X_\mathrm{train} = \{S^{(k)}\}_{k=1}^{N_\mathrm{train}}$ contains atomic structures and $Y_\mathrm{train} = \{E_k^\mathrm{ref}, \{\mathbf{F}_{u,k}^\mathrm{ref}\}_{u=1}^{N_\mathrm{at}^{(k)}}\}_{k=1}^{N_\mathrm{train}}$ provides the corresponding reference energies and forces. The loss function is defined as
\begin{equation}
    \label{eq:loss}
    \begin{split}
        \mathcal{L}\left( \boldsymbol{\theta}, D_\mathrm{train}\right) = \sum_{k=1}^{N_\mathrm{train}} \Bigg[ & C_\mathrm{e}^{(k)} \Big\lVert E_k^\mathrm{ref} - E(S^{(k)}, \boldsymbol{\theta})\Big\rVert_2^2 \\ + & C_\mathrm{f} \sum_{u=1}^{N_\mathrm{at}^{(k)}} \Big\lVert \mathbf{F}_{u,k}^\mathrm{ref} - \mathbf{F}_u\left(S^{(k)}, \boldsymbol{\theta}\right)\Big\rVert_2^2 \Bigg],
    \end{split}
\end{equation}
where $E(S^{(k)}, \boldsymbol{\theta})$ and $\mathbf{F}_u(S^{(k)}, \boldsymbol{\theta})$ are energies and forces predicted by the ICTP models, including the contributions from analytic pair potentials. To balance the relative contributions of energies and forces, we set $C_\mathrm{e}^{(k)}=1 / N_\mathrm{at}^{(k)}$ and $C_\mathrm{f} = 0.05$~\AA$^2$ during training.

\subsection*{Long-range dispersion and electrostatics}

Message-passing architectures can capture interactions beyond the local cutoff radius $r_\mathrm{c}$, set to 5~\AA{} in this work. For example, two message-passing layers yield an effective interaction range of $r_\mathrm{c}^\mathrm{eff} = 10~\AA$. However, increasing the number of layers leads to significant computational cost and does not guarantee that learned long-range interactions follow the correct power-law decay. Therefore, we incorporate analytic long-range corrections (see \figref{fig:overview}~(c)) into the energy in \eqref{eq:energy-model}, which we denote as $E_\mathrm{local}$, and define the total energy as
\begin{equation}
    \label{eq:local-es-disp}
    E = E_\mathrm{local} + E_\mathrm{disp} + E_\mathrm{es}.
\end{equation}
Here, $E_\mathrm{disp} = E_\mathrm{disp}\left(S, \mathbf{q}\left(S, \boldsymbol{\theta}\right)\right)$ and $E_\mathrm{es} = E_\mathrm{es}\left(S, \mathbf{q}\left(S, \boldsymbol{\theta}\right)\right)$ are dispersion and electrostatic contributions that depend on machine-learned partial charges $\mathbf{q}\left(S, \boldsymbol{\theta}\right) = \left\{q_u(S_u, \boldsymbol{\theta})\right\}_{u=1}^{N_\mathrm{at}}$.

Dispersion is modeled using the two-body term of the D4 correction,\cite{Caldeweyher2019} with pairwise coefficients dependent on the learned partial charges. Following prior work,\cite{Unke2021b} we treat parameters that vary between density functionals as learnable and introduce a learnable scaling factor for the tabulated reference charges. Due to the fast decay of two-body dispersion interactions, we truncate them at a cutoff radius of 9~\AA{}. To ensure that the energy and the forces vanish at the cutoff, we apply the shifted force correction.\cite{Fennell2006} Additionally, we interpolate the coordination number to zero at the cutoff using a switch function.

Electrostatic interactions are modeled using the Coulomb potential for isolated systems during training, and either Ewald summation or the SPME method\cite{Essmann1995} for periodic systems during inference. SPME efficiently evaluates the reciprocal-space contribution, allowing us to exploit the automatic differentiation capabilities of PyTorch.\cite{Christiansen2025} A real-space cutoff of 9~\AA{} is used throughout this work, while the reciprocal-space cutoff and smearing parameter are defined individually for each simulated system following established practices.\cite{Eastman2023} We used an error tolerance of $5 \times 10^{-5}$ and a B-spline interpolation of order 5 for the reciprocal-space evaluation.

The dispersion and electrostatic energy terms defined in this work exclude interactions between atom pairs within the short-range cutoff of 5~\AA, as the local energy model already accounts for these interactions. To ensure consistency and transferability between the Coulomb potential used during training and the Ewald or SPME methods used during inference, total charge conservation is enforced by uniformly redistributing any residual net charge across all atoms.\cite{Unke2021b, Kabylda2025}

Unlike prior work,\cite{Unke2021b, Unke2024, Anstine2025, Kabylda2025} we do not constrain partial charges by including their reference values or dipole moments in the training loss. Instead, we treat partial charges as latent variables, similar to atomic energies, and train them exclusively from reference energies and forces. Our approach is related to Ref.~\citenum{Cheng2025}, but it avoids using periodic boxes for isolated systems, thereby mitigating inductive biases that may hinder transferability to bulk systems.

\subsection*{Short-range repulsion}

To ensure correct asymptotic behavior as $r \rightarrow 0$ and aid the training process, we include a short-range repulsion term $E_\mathrm{rep}$ in the total energy (see \figref{fig:overview}~(c))
\begin{equation}
    \label{eq:rep-local-es-disp}
    E = E_\mathrm{rep} + E_\mathrm{local} + E_\mathrm{es} + E_\mathrm{disp}.
\end{equation}
In particular, we use the Ziegler–Biersack–Littmark (ZBL) potential,\cite{Ziegler1985} with all parameters treated as learnable and initialized from tabulated values. A smooth cutoff function is applied to the pairwise contributions, with element-specific radii determined by the sum of covalent radii.

\subsection*{Training and test datasets}

We train ICTP models on an expanded and curated version of the SPICE-v2 dataset,\cite{Eastman2023b, Eastman2024} comprising 2,008,628 molecules and molecular clusters with reference energies and forces computed at the $\omega$B97M-D3(BJ)/def2-TZVPPD level of theory.\cite{Najibi2018, Weigend05, Rappoport2010, Grimme2010, Grimme2011} All DFT calculations were performed using the PSI4 software package.\cite{Smith2020psi4} Following prior work,\cite{Kovacs2025} we augmented SPICE-v2 with additional molecules from the QMugs dataset\cite{Isert2022} (50--90 atoms) and water clusters\cite{Schran2021} (up to 150 atoms). We further included 1092 \ce{NaCl}-water clusters, generated by substituting water molecules in SPICE-v2 and the added water clusters with \ce{Na+} and \ce{Cl-} ion pairs, ensuring charge neutrality. All additional reference energies and forces were computed at the same level of theory to maintain consistency across the dataset.

To identify and remove outliers from the original SPICE-v2 dataset, we trained three medium-sized ICTP-LR models and used them to detect structures with the highest prediction errors in energies and forces. Structures that consistently exhibited large errors across all three models were re-evaluated using the same level of theory. Any structures for which the new DFT calculations failed to converge or remained inconsistent with model predictions were excluded. In total, 890 outlier structures were removed from the dataset.

The final dataset is split such that 95~\% of structures were used for training and validation, while the remaining 5~\% were held out for testing. This split was performed at the subset level, meaning that conformers of the same molecule may appear in both the training/validation and test datasets. Therefore, we treated the held-out test datasets as in-distribution. To evaluate our models on data not seen during training, we used separate test-only datasets,\cite{Eastman2024} along with two torsion datasets.\cite{Lahey2020, Rai2022} For the latter, we recomputed reference energies and forces to ensure consistency with the chosen level of theory.

\subsection*{Evaluation with classical force fields}

Molecules from the test-only datasets were assigned GAFF2 parameters\cite{Wang2004} and AM1-BCC\cite{Jakalian2000,Jakalian2002} partial charges using AmberTools23,\cite{Case2023} via the \texttt{antechamber} package. Energies and forces were computed using OpenMM version 8.2.\cite{Eastman2023} Molecules containing additional fragments, such as cofactors or metal ions, were excluded from these calculations, as \texttt{antechamber} supports only single molecular entities.

\subsection*{System preparation}

In this work, we performed simulations of Ala3 in blocked and cationic forms, the mini-protein Trp-Cage (PDB ID: 1L2Y)\cite{Neidigh2002}, and Crambin (PDB ID: 1EJG).\cite{Jelsch2000} Ala3 systems were built using the \texttt{tleap} program from AmberTools23.\cite{Case2023} For the blocked form of Ala3, the \ce{N}- and \ce{C}-termini were capped with an acetyl group (ACE) and a \ce{N}-methyl amide group (NME), respectively. In the cationic form, we used a protonated amine group (\ce{NH$_3^{+}$}) at the \ce{N}-terminus and an NHE group at the \ce{C}-terminus, which corresponds to an amide termination (\ce{C(=O)-NH2}).

All peptide and protein structures were solvated in periodic TIP3P water boxes using AmberTools23 and neutralized with \ce{Na+} and \ce{Cl-} ions. Solvation boxes were prepared to ensure a minimum distance of 10~\AA{} between the solute and the box edges. The resulting system sizes are 2817 atoms (blocked Ala3), 2754 atoms (cationic Ala3), 6441 atoms (Trp-cage), and 10,933 atoms (Crambin).

We generated water boxes from geometry-optimized TIP3P water molecules. We built \ce{NaCl}-water mixtures by randomly replacing water molecules with \ce{Na+} and \ce{Cl-} ion pairs to achieve target molalities. The pure water box contains 3072 atoms, while the \ce{NaCl}-water mixtures include 2598 atoms (0.99~mol/kg), 2476 atoms (1.99~mol/kg), 2420 atoms (3.07~mol/kg), 2372 atoms (4.05~mol/kg), and 2328 atoms (5.0~mol/kg).

\subsection*{Simulation details}

All simulations in this work were performed using the differentiable molecular simulation framework (DIMOS).\cite{Christiansen2025} For classical FFs, a 9~\AA{} cutoff was applied to the Lennard--Jones potential, with a switching function smoothly reducing interactions to zero between 7.5~\AA{} and 9~\AA. A dispersion correction was also included. Electrostatic interactions were treated using the SPME method with the 9~\AA{} real-space cutoff. The reciprocal-space cutoff and smearing parameter were chosen following established practices,\cite{Eastman2023} with an error tolerance of $5 \times 10^{-5}$ and a B-spline interpolation of order 5.

All MD simulations were performed in the isothermal-isobaric ($NpT$) ensemble. A Langevin thermostat with a friction coefficient of 0.01~fs$^{-1}$ was used to control the temperature, while an isotropic Monte Carlo barostat was applied every 100 steps to maintain a pressure of 1.0~bar. We set, unless stated otherwise, the integration time step to 0.5~fs, and the first 0.2~ns of each trajectory were used for equilibration. All simulations were initialized using structures prepared as described in the previous section.

For pure liquid water, simulations were conducted over a temperature range of 273.15--373.15~K in 25~K increments. For the \ce{NaCl}-water mixture, salt concentrations were varied across 0.99, 1.99, 3.07, 4.05, and 5.0~mol/kg, with all simulations performed at 298.15~K. To compute densities and RDFs of pure liquid water and the \ce{NaCl}-water mixture, frames were recorded every 200 steps over 1.2~ns trajectories. RDFs were evaluated using MDAnalysis.\cite{Michaud2011, Gowers2016} For Crambin, molecular dynamics simulations were also performed for 1.2~ns at 298.15~K, with every frame stored to enable the calculation of vibrational power spectra at a resolution of 0.5~fs. The power spectrum was computed as the Fourier transform of the velocity autocorrelation function using the Travis program with default parameters.\cite{Brehm2011, Brehm2020}

For Ala3, we used PLUMED to perform multiple-walker, well-tempered metadynamics simulations\cite{Raiteri2006, Barducci2008, Massimiliano2009, Tribello2014} biasing the central $\phi$ and $\psi$ backbone dihedral angles. A 0.2~ns equilibration at 298.15~K was performed prior to the metadynamics runs. Six independent walkers were then initiated from the same equilibrated structure using different random seeds, each run for 10~ns, yielding a combined total of 60~ns of sampling. The parameters of the Gaussian bias potential in the metadynamics were a rate of deposition of 1000 steps, a Gaussian height of 0.2~kcal/mol ($\approx$8.67~meV), a Gaussian width of 0.35~rad for each of the collective variables, and a bias factor of 6. The frames were recorded every 1000 steps. Final FES were obtained by reweighting the trajectories using the bias potential obtained at the end of the simulation, assuming a constant bias during the simulation.\cite{Branduardi2012}

For Trp-cage, we used PLUMED to perform multiple-walker, parallel-bias,\cite{Pfaendtner2015} well-tempered metadynamics simulations, preceded by a 0.2~ns equilibration at 298.15~K. Six walkers were used, each run for 10~ns. Following previous work,\cite{Pfaendtner2015} we biased six collective variables that capture key aspects of protein conformational dynamics; see \figref{fig:trp-crambin-results}. The integration time step was set to 1.0~fs. Gaussian hills were deposited every 500 steps with an initial height of 0.25~kcal/mol ($\approx$10.84~meV). Gaussian widths were set to 0.1~\AA, 0.6, 0.3, 0.4, 0.3, and 0.6, respectively. A bias factor of 8 was used. Frames were recorded every 500 steps. Final FES were obtained by reweighting the trajectories.\cite{Branduardi2012}

\subsection*{Normalized error}

The normalized error (NE) in \figref{fig:inference-results}~(a) is calculated as
\begin{equation}
    \label{eq:norm_error}
    \text{NE} = \frac{1}{2} \left(\frac{\text{F-RMSE}}{\text{max}\,\text{F-RMSE}} + \frac{\text{E-RMSE}}{\text{max}\,\text{E-RMSE}}\right),
\end{equation}
where E-RMSE and F-RMSE are relative energy and force errors, respectively.

\section*{Data availability}

The training and test datasets, as well as the trained models, will be made available at \url{https://10.5281/zenodo.16607765}.

\section*{Code availability}

The source code used in this study is available on GitHub at \url{https://github.com/nec-research/ictp}.

\section*{Acknowledgements}

The authors thank David Holzm{\"u}ller, Federico Errica, Germ{\'a}n Molpeceres, Henrik Christiansen, Makoto Takamoto, and Takashi Maruyama for useful discussions. Mathias Niepert acknowledges support from the Deutsche Forschungsgemeinschaft (DFG, German Research Foundation) under Germany's Excellence Strategy - EXC 2075 – 390740016 and the Stuttgart Center for Simulation Science (SimTech).

\section*{Author contributions}

Viktor Zaverkin: Conceptualization, Data curation, Formal analysis, Investigation, Methodology, Software, Visualization, Writing -- original draft, Writing -- review \& editing; Matheus Ferraz: Conceptualization, Data curation, Writing -- review \& editing; Francesco Alesiani: Conceptualization, Writing -- review \& editing; Mathias Niepert: Conceptualization, Writing -- review \& editing.

\section*{Competing interests}

The authors declare no competing interests.

\section*{Additional information}

\textbf{Supplementary Information} accompanies.

\bibliography{references}

%%%%%%%%%%%%%%%%%%%%%%%%%%%%%%%%%%%%%%%%%%%%%%%%%%%%%%%%%%%%

\clearpage

\begin{appendices}

\setcounter{equation}{0}
\setcounter{figure}{0}
\setcounter{table}{0}

\renewcommand{\theequation}{\arabic{equation}}

\renewcommand{\figurename}{Supplementary Figure}
\renewcommand{\tablename}{Supplementary Table}

\renewcommand{\figref}[1]{Supplementary \fig~\ref{#1}}
\renewcommand{\Figref}[1]{Supplementary \Fig~\ref{#1}}
\renewcommand{\figvref}[1]{Supplementary \fig~\vref{#1}}
\renewcommand{\Figvref}[1]{Supplementary \Fig~\vref{#1}}

\renewcommand{\tabref}[1]{Supplementary Table~\ref{#1}}
\renewcommand{\Tabref}[1]{Supplementary Table~\ref{#1}}
\renewcommand{\tabvref}[1]{Supplementary Table~\vref{#1}}
\renewcommand{\Tabvref}[1]{Supplementary Table~\vref{#1}}

\renewcommand{\eqref}[1]{Supplementary Equation~(\ref{#1})}
\renewcommand{\Eqref}[1]{Supplementary Equation~(\ref{#1})}
\renewcommand{\eqvref}[1]{Supplementary Equation~(\ref{#1}) \vpageref{#1}}
\renewcommand{\Eqvref}[1]{Supplementary Equation~(\ref{#1}) \vpageref{#1}}

\renewcommand{\secref}[1]{Supplementary Section~\ref{#1}}
\renewcommand{\Secref}[1]{Supplementary Section~\ref{#1}}
\renewcommand{\secvref}[1]{Supplementary Section~\ref{#1} \vpageref{#1}}
\renewcommand{\Secvref}[1]{Supplementary Section~\ref{#1} \vpageref{#1}}

\section*{Supplementary results}

\subsection*{Accuracy for the in- and out-of-distribution datasets}

We report RMSEs and MAEs in the predicted energies, atomic forces, and intermolecular forces across all datasets used in this work. \Tabref{tab:test_errors-rmse} and \tabref{tab:test_errors-mae} summarize the RMSEs and MAEs for the GAFF2, the short-ranged ICTP-SR(M) model, and the ICTP-LR models with explicit long-range dispersion and electrostatics. \Tabref{tab:comparison-allegro} and \tabref{tab:comparison-mace} further compare the energy and force MAEs of the ICTP models to those of Allegro and MACE-OFF24(M).

\subsection*{Bulk water with various salt concentrations}

We report relative errors in the predicted density of pure liquid water and \ce{NaCl}-water mixtures as a function of temperature and salt concentration. \Tabref{tab:density-errors} and \tabref{tab:density-errors-nacl} summarize these errors for the TIP3P and AMBER14SB+TIP3P classical FFs, the short-ranged ICTP-SR(M) model, and the ICTP-LR models with explicit long-range dispersion and electrostatics.

\subsection*{Alanine tripeptide in aqueous solutions}

\begin{figure*}[t!]
	\includegraphics[width=\linewidth]{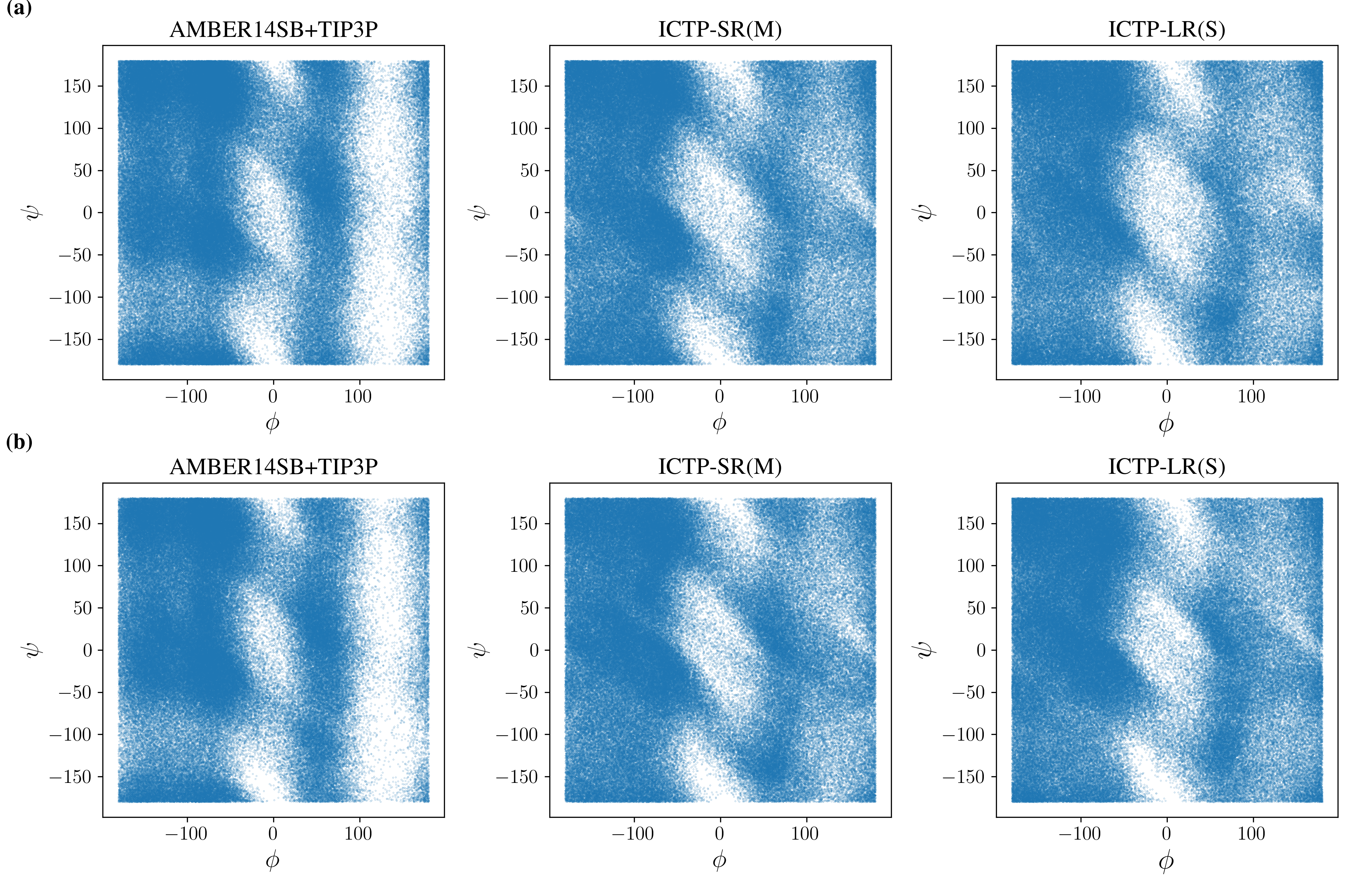}
	\caption{Ramachandran plots of the central backbone dihedral angles ($\phi$ and $\psi$) for Ala3 in its (a) cationic and (b) blocked forms. Results are obtained from metadynamics simulations using the AMBER14SB+TIP3P classical force field, the short-range ICTP-SR(M) model, and the long-range ICTP-LR(S) model.}
	\label{fig:ala3-ramachandran}
\end{figure*}

\Tabref{tab:ala3_conformations} presents the backbone dihedral angles ($\phi$ and $\psi$) and relative free energies of representative low-energy conformers obtained from metadynamics simulations using AMBER14SB+TIP3P and the ICTP models. \Figref{fig:ala3-ramachandran} demonstrates the corresponding Ramachandran plots. We also report $J$-coupling constants derived from the dihedral angle distributions in \tabref{tab:ala3_jcoupling}.

\subsection*{Small proteins---Trp-cage and Crambin}

\begin{figure*}[t!]
	\includegraphics[width=\linewidth]{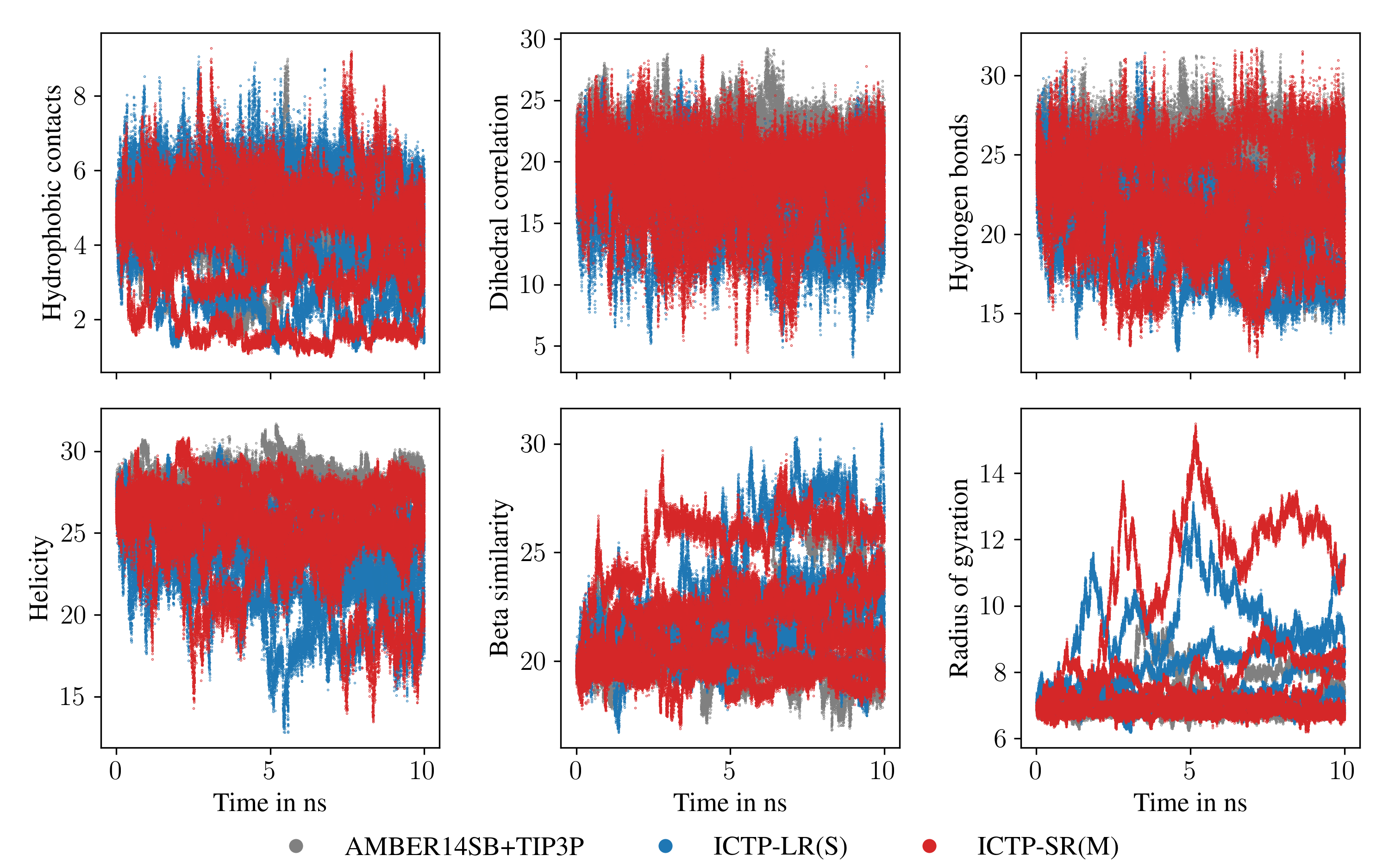}
	\caption{Time evolution of six CVs used to describe the conformational dynamics of Trp-cage. The CVs include the number of C$_\gamma$-hydrophobic contacts, the dihedral correlation, the number of backbone hydrogen bonds, the $\alpha$- and $\beta$-dihedral fractions, and the C$_\alpha$-radius of gyration. Each time series represents a separate walker from metadynamics simulations, with 10~ns of trajectory shown per walker.}
	\label{fig:trp-cvs}
\end{figure*}

\begin{figure*}[t!]
	\includegraphics[width=\linewidth]{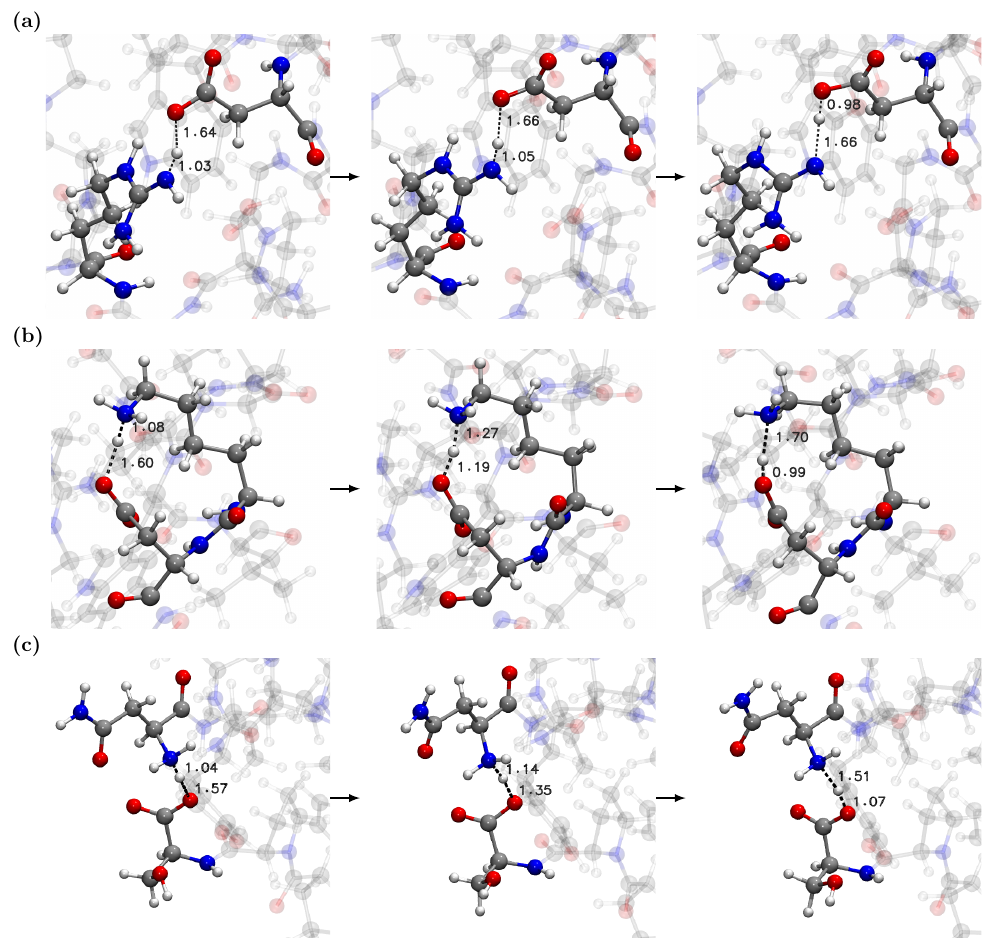}
	\caption{Representative snapshots of proton transfer reactions observed in Trp-cage during metadynamics simulations with the ICTP models. (a) Proton transfer between an $\ce{=NH2^+}$ group on R16 and a nearby $\ce{-COO^-}$ group on D9, observed exclusively with ICTP-LR(S), including the reverse reaction. (b) Proton transfer between an $\ce{-NH3^+}$ group on K8 and a $\ce{-COO^-}$ group on D9, also observed with ICTP-LR(S). (c) Terminal $\ce{-NH3^+}$ to $\ce{-COO^-}$ proton transfer involving N1 and S20, observed with ICTP-SR(M). All distances are in \AA.}
	\label{fig:proton_transfer}
\end{figure*}

\Figref{fig:trp-cvs} shows the time evolution of all six CVs across independent walkers, each spanning 10~ns of metadynamics simulation. \Figref{fig:proton_transfer} demonstrates representative snapshots of the observed proton transfer events.

\section*{Supplementary methods}

\subsection*{Training and test datasets}

\Tabref{tab:datasets} provides an overview of all datasets used in this work, including the number of structures in the training/validation and held-out test splits, typical molecular sizes, and the corresponding atomic species. The test-only datasets were not included during training and serve exclusively for benchmarking model generalization.

\begin{table*}[t!]
    \caption{RMSEs in the predicted energies/atomic forces/intermolecular forces for the individual test datasets. Results are provided for the GAFF2, the short-ranged ICTP-SR(M) model, as well as for the ICTP-LR(S), ICTP-LR(M), ICTP-LR(M)$^\ast$, and ICTP-LR(L) models, which explicitly model long-range dispersion and electrostatics. The ICTP-LR(M)$^\ast$ model corresponds to ICTP-LR(M) trained without the \ce{NaCl}-Water Clusters dataset. Energy RMSEs are given in meV/atom, while errors in the predicted atomic and intermolecular forces are given in meV/\AA. \label{tab:test_errors-rmse}}
    \begin{center}
        \resizebox{\textwidth}{!}{
        \begin{tabular}{lllllll}
            \toprule
            Dataset                                     & GAFF2             & ICTP-LR(S)            & ICTP-LR(M)        & ICTP-LR(M)$^\ast$     & ICTP-LR(L)        & ICTP-SR(M)            \\
            \cmidrule(lr){1-1} \cmidrule(lr){2-7}
            Solvated Amino Acids                        & --                & 1.59/45.48/19.34      & 1.04/33.11/14.02  & 1.22/34.20/14.51      & 0.82/26.38/11.14  & 1.50/38.34/17.51      \\
            Dipeptides                                  & --                & 1.93/44.99/0.13       & 1.29/33.80/0.13   & 1.27/33.56/0.13       & 0.98/25.60/0.13   & 1.89/35.16/0.13       \\
            DES370K Monomers                            & --                & 2.63/34.50/0.45       & 2.04/25.12/0.45   & 1.93/25.16/0.45       & 1.41/18.43/0.45   & 3.64/25.16/0.45       \\
            DES370K Dimers                              & --                & 3.95/38.94/11.40      & 3.17/28.19/9.18   & 2.81/28.23/9.35       & 2.07/21.00/6.94   & 3.63/29.43/9.89       \\
            PubChem                                     & --                & 5.04/69.17/0.38       & 3.34/52.34/0.29   & 3.35/52.39/0.49       & 2.46/40.68/0.32   & 3.56/52.81/0.31       \\
            Ion Pairs\textsuperscript{\emph{a}}         & --                & 24.00/123.38/123.38   & 32.97/79.09/79.09 & 15.61/80.46/80.46     & 22.21/59.29/59.29 & 143.38/451.87/451.87  \\
            Solvated PubChem                            & --                & 3.77/83.57/22.84      & 2.90/66.58/18.37  & 3.04/68.19/18.63      & 2.39/56.20/15.39  & 3.07/69.12/20.75      \\
            Water Clusters                              & --                & 1.25/28.68/21.18      & 0.76/20.53/14.67  & 0.96/22.19/15.35      & 0.82/15.58/11.28  & 1.13/24.88/17.96      \\
            Amino Acid Ligand                           & --                & 2.55/43.18/4.93       & 2.00/32.87/5.12   & 2.03/32.95/3.71       & 1.68/26.45/3.91   & 2.58/35.52/6.02       \\
            QMugs                                       & --                & 1.57/72.62/0.22       & 1.15/55.77/0.22   & 1.20/55.49/0.22       & 0.94/44.01/0.22   & 1.29/55.55/0.22       \\
            NaCl-Water Clusters                         & --                & 3.29/50.89/32.92      & 2.11/38.12/24.39  & 12.86/202.33/103.64   & 1.74/30.04/20.21  & 4.57/53.68/37.99      \\
            \cmidrule(lr){1-1} \cmidrule(lr){2-7}
            \multicolumn{7}{l}{Test-only datasets}                                                                                                                                          \\
            \cmidrule(lr){1-1} \cmidrule(lr){2-7}
            Large Ligands\textsuperscript{\emph{b}}     & NA/781.09/0.28    & 2.51/103.59/0.17      & 3.03/104.85/0.17  & 3.07/121.65/0.17      & 2.89/105.89/0.17  & 4.33/121.96/0.17      \\
            Small Ligands\textsuperscript{\emph{b}}     & NA/836.66/0.43    & 7.41/95.41/1.32       & 7.68/87.68/0.75   & 7.72/90.41/1.14       & 7.90/80.81/0.92   & 7.83/88.24/0.88       \\
            Pentapeptides                               & NA/548.96/0.19    & 2.12/52.76/0.10       & 1.85/41.11/0.10   & 1.83/41.04/0.10       & 1.57/32.84/0.10   & 4.92/44.31/0.10       \\
            TorsionNet-500                              & NA/475.81/0.33    & 2.63/37.30/0.22       & 1.65/27.68/0.22   & 1.63/27.86/0.22       & 1.15/21.20/0.22   & 1.64/28.09/0.22       \\
            Biaryl Torsion                              & NA/498.41/0.32    & 2.86/29.81/0.11       & 1.66/21.03/0.11   & 1.44/21.17/0.11       & 1.13/16.15/0.11   & 1.56/21.02/0.11       \\
            \bottomrule
        \end{tabular}}
    \end{center}
    \footnotesize{\textsuperscript{\emph{a}} For the Ion Pairs dataset, we report force RMSEs only for the $z$-component, as the $x$ and $y$ components are zero.}\\
    \footnotesize{\textsuperscript{\emph{b}} A few outliers impact the reported errors for the Large Ligands and Small Ligands datasets. Replacing RMSEs with the more robust MAEs reduces the values to 1.47/38.08/0.08 and 2.06/37.24/0.13 for the ICTP-LR(M) model, respectively. For more details, we refer to \tabref{tab:test_errors-mae}.}
\end{table*}

\begin{table*}[t!]
    \caption{MAEs in the predicted energies/atomic forces/intermolecular forces for the individual test datasets. Results are provided for the GAFF2, the short-ranged ICTP-SR(M) model, as well as for the ICTP-LR(S), ICTP-LR(M), ICTP-LR(M)$^\ast$, and ICTP-LR(L) models, which explicitly model long-range dispersion and electrostatics. The ICTP-LR(M)$^\ast$ model corresponds to ICTP-LR(M) trained without the \ce{NaCl}-Water Clusters dataset. Energy MAEs are given in meV/atom, while errors in the predicted atomic and intermolecular forces are given in meV/\AA. \label{tab:test_errors-mae}}
    \begin{center}
    \resizebox{\textwidth}{!}{
        \begin{tabular}{lllllll}
            \toprule
            Dataset                & GAFF2             & ICTP-LR(S)        & ICTP-LR(M)        & ICTP-LR(M)$^\ast$ & ICTP-LR(L)        & ICTP-SR(M)            \\
            \midrule
            Solvated Amino Acids    & --                & 1.32/31.31/12.20  & 0.78/23.21/8.93   & 0.96/23.93/9.17   & 0.65/18.48/7.06   & 1.18/27.10/11.04      \\
            Dipeptides              & --                & 1.36/30.75/0.08   & 0.91/22.91/0.08   & 0.89/22.84/0.08   & 0.71/17.28/0.08   & 1.32/23.78/0.08       \\
            DES370K Monomers        & --                & 1.85/22.52/0.17   & 1.35/16.31/0.17   & 1.33/16.32/0.17   & 0.95/11.88/0.17   & 1.87/16.34/0.17       \\
            DES370K Dimers          & --                & 2.17/21.42/3.46   & 1.51/15.24/2.69   & 1.45/15.27/2.62   & 1.01/11.06/2.08   & 1.70/15.94/3.00       \\
            PubChem                 & --                & 2.56/40.92/0.13   & 1.74/30.55/0.13   & 1.72/30.67/0.13   & 1.23/23.30/0.13   & 1.78/30.84/0.13       \\
            Ion Pairs               & --                & 19.81/76.45/76.45 & 26.39/59.93/59.93 & 12.12/54.17/54.17 & 17.00/44.80/44.80 & 114.12/355.50/355.50  \\
            Solvated PubChem        & --                & 2.13/40.78/13.01  & 1.58/31.63/10.21  & 1.60/32.07/10.37  & 1.19/25.88/8.45   & 1.77/34.46/11.81      \\
            Water Clusters          & --                & 1.04/21.84/15.11  & 0.57/15.44/10.34  & 0.75/16.75/10.92  & 0.64/11.78/7.94   & 0.89/18.74/12.75      \\
            Amino Acid Ligand       & --                & 1.43/23.91/1.38   & 1.02/17.39/1.08   & 1.02/17.51/1.08   & 0.77/13.01/0.89   & 1.19/18.80/1.21       \\
            QMugs                   & --                & 1.16/47.16/0.09   & 0.83/36.24/0.09   & 0.88/35.80/0.09   & 0.70/28.42/0.09   & 0.96/36.11/0.09       \\
            NaCl Water Clusters     & --                & 2.65/36.53/21.81  & 1.72/27.07/16.20  & 9.23/132.79/67.35 & 1.43/21.04/12.83  & 3.57/38.03/24.22      \\
            \midrule
            \multicolumn{7}{l}{Test-only datasets}                                                                                                             \\
            \midrule
            Large Ligands           & NA/405.78/0.15    & 1.56/47.04/0.08   & 1.47/38.08/0.08   & 1.38/38.25/0.08   & 1.15/31.35/0.08   & 1.48/38.50/0.08       \\
            Small Ligands           & NA/427.26/0.22    & 2.32/46.92/0.13   & 2.06/37.24/0.13   & 1.97/37.47/0.13   & 1.75/29.93/0.13   & 2.09/37.45/0.13       \\
            Pentapeptides           & NA/362.78/0.11    & 1.59/36.51/0.06   & 1.34/28.42/0.06   & 1.39/28.38/0.06   & 1.22/22.62/0.06   & 2.08/30.14/0.06       \\
            TorsionNet-500          & NA/298.70/0.17    & 1.96/23.54/0.09   & 1.14/17.22/0.09   & 1.14/17.22/0.09   & 0.82/13.06/0.09   & 1.13/17.23/0.09       \\
            Biaryl Torsion          & NA/316.75/0.18    & 2.37/19.02/0.07   & 1.20/13.53/0.07   & 1.04/13.70/0.07   & 0.84/10.13/0.07   & 1.16/13.56/0.07       \\
            \bottomrule
        \end{tabular}
    }
    \end{center}
    \footnotesize{\textsuperscript{\emph{a}} For the Ion Pairs dataset, we report force MAEs only for the $z$-component, as the $x$ and $y$ components are zero.}\\
\end{table*}

\begin{table*}[t!]
    \caption{Comparison of MAEs in the predicted energies/atomic forces for the test-only datasets between the ICTP and Allegro models.\textsuperscript{\emph{a}} Energy MAEs are given in meV, while errors in the predicted forces are given in meV/\AA. \label{tab:comparison-allegro}}
    \begin{center}
    \resizebox{\textwidth}{!}{
        \begin{tabular}{llllllllllll}
            \toprule
            Dataset                & Allegro-U(S)  & Allegro-U(M)  & Allegro-U(L)  & Allegro-NTC(S)    & Allegro-NTC(M)    & Allegro-NTC(L)    & ICTP-LR(S)    & ICTP-LR(M)    & ICTP-LR(M)$^\ast$ & ICTP-LR(L)    & ICTP-SR(M)    \\
            \cmidrule(lr){1-1} \cmidrule(lr){2-4} \cmidrule(lr){5-7} \cmidrule(lr){8-12}
            Large Ligands           & 220/42        & 184/35        & 165/27        & 102/40            & 75/32             & 57/25             & 116.12/47.04  & 109.57/38.08  & 103.15/38.25      & 86.24/31.35   & 110.00/38.50  \\
            Small Ligands           & 315/43        & 298/36        & 279/28        & 74/39             & 56/31             & 42/23             & 102.30/46.92  & 90.30/37.24   & 86.36/37.47       & 76.45/29.93   & 91.99/37.45   \\
            Pentapeptides           & 2164/47       & 2173/43       & 2231/36       & 83/35             & 74/29             & 57/23             & 143.50/36.51  & 120.95/28.42  & 125.92/28.38      & 111.34/22.62  & 180.97/30.14  \\
            \bottomrule
        \end{tabular}
    }
    \end{center}
    \footnotesize{\textsuperscript{\emph{a}} The Allegro models are trained on a subset of the SPICE-v2 dataset limited to systems with a neutral total charge. The results are for the original unrestricted test-only datasets (U) or a subset containing systems with neutral total charge (NTC).}
\end{table*}

\begin{table*}[t!]
    \caption{Comparison of MAEs in the predicted energies/atomic forces for the held-out test datasets between the ICTP models and MACE-OFF24(M).\textsuperscript{\emph{a}} Energy MAEs are given in meV/atom, while errors in the predicted forces are given in meV/\AA. \label{tab:comparison-mace}}
    \begin{center}
    \resizebox{\textwidth}{!}{
        \begin{tabular}{lllllll}
            \toprule
            Dataset                & MACE-OFF24(M) & ICTP-LR(S)  & ICTP-LR(M)   & ICTP-LR(M)$^\ast$    & ICTP-LR(L)   & ICTP-SR(M)     \\
            \cmidrule(lr){1-1} \cmidrule(lr){2-2} \cmidrule(lr){3-7}
            Solvated Amino Acids    & 1.3/23.2      & 1.32/31.31  & 0.78/23.21   & 0.96/23.93           & 0.65/18.48   & 1.18/27.10     \\
            Dipeptides              & 0.5/14.4      & 1.36/30.75  & 0.91/22.91   & 0.89/22.84           & 0.71/17.28   & 1.32/23.78     \\
            DES370K Monomers        & 0.6/9.6       & 1.85/22.52  & 1.35/16.31   & 1.33/16.32           & 0.95/11.88   & 1.87/16.34     \\
            DES370K Dimers          & 0.6/9.3       & 2.17/21.42  & 1.51/15.24   & 1.45/15.27           & 1.01/11.06   & 1.70/15.94     \\
            PubChem                 & 1.0/22.1      & 2.56/40.92  & 1.74/30.55   & 1.72/30.67           & 1.23/23.30   & 1.78/30.84     \\
            Solvated PubChem        & 1.2/22.8      & 2.13/40.78  & 1.58/31.63   & 1.60/32.07           & 1.19/25.88   & 1.77/34.46     \\
            Amino Acid Ligand       & 1.5/24.2      & 1.43/23.91  & 1.02/17.39   & 1.02/17.51           & 0.77/13.01   & 1.19/18.80     \\
            QMugs                   & 0.8/23.8      &1.16/47.16   & 0.83/36.24   & 0.88/35.80           & 0.70/28.42   & 0.96/36.11     \\
            \bottomrule
        \end{tabular}
    }
    \end{center}
    \footnotesize{\textsuperscript{\emph{a}} MACE-OFF24(M) is trained and tested on a subset of the SPICE-v2 dataset that includes 10 chemical elements (H, C, N, O, F, P, S, Cl, Br, and I) and has a neutral formal charge.}
\end{table*}

\begin{table*}[t!]
    \caption{Relative errors in the predicted density of pure liquid water as a function of temperature. Results are provided for the TIP3P classical FF, the short-ranged ICTP-SR(M) model, as well as for the ICTP-LR(S), ICTP-LR(M), ICTP-LR(M)$^\ast$, and ICTP-LR(L) models, which explicitly model long-range dispersion and electrostatics. The ICTP-LR(M)$^\ast$ model corresponds to ICTP-LR(M) trained without the \ce{NaCl}-Water Clusters dataset. All errors are reported as percentages. \label{tab:density-errors}}
    \begin{center}
        \begin{tabular}{llllllllllll}
            \toprule
            Temperature     & TIP3P     & ICTP-LR(S)    & ICTP-LR(M)    & ICTP-LR(M)$^\ast$ & ICTP-LR(L)    & ICTP-SR(M)    \\
            \midrule
            273.15          & 2.83      & 3.90          & 4.33          & 2.26              & 3.83          & 4.11          \\
            298.15          & 1.32      & 3.47          & 2.96          & 1.80              & 3.03          & 3.10          \\
            323.15          & 0.17      & 3.52          & 2.12          & 1.38              & 2.33          & 2.31          \\
            348.15          & 0.82      & 3.36          & 1.67          & 1.24              & 2.17          & 1.78          \\
            373.15          & 2.13      & 3.31          & 1.39          & 1.20              & 1.84          & 1.15          \\
            \midrule
            Avg.            & 1.45      & 3.51          & 2.50          & 1.58              & 2.64          & 2.49          \\
            \bottomrule
        \end{tabular}
    \end{center}
\end{table*}

\begin{table*}[t!]
    \caption{Relative errors in the predicted density of \ce{NaCl}-water mixtures as a function of the \ce{NaCl} concentration. Results are provided for the TIP3P classical FF, the short-ranged ICTP-SR(M) model, as well as for the ICTP-LR(S), ICTP-LR(M), ICTP-LR(M)$^\ast$, and ICTP-LR(L) models, which explicitly model long-range electrostatics. The ICTP-LR(M)$^\ast$ model corresponds to ICTP-LR(M) trained without the \ce{NaCl}-Water Clusters dataset.\textsuperscript{\emph{a}} All errors are reported as percentages. \label{tab:density-errors-nacl}}
    \begin{center}
    \resizebox{\textwidth}{!}{
        \begin{tabular}{llllllllllll}
            \toprule
            Concentration   & AMBER14SB+TIP3P   & ICTP-LR(S)    & ICTP-LR(M)    & ICTP-LR(M)$^\ast$ & ICTP-LR(L)    & ICTP-SR(M)    \\
            \midrule
            0.99            & 1.72              & 1.26          & 3.59          & NA                & 4.60          & 3.74          \\
            1.99            & 1.57              & 0.36          & 4.23          & NA                & 5.51          & 3.84          \\
            3.07            & 1.60              & 2.00          & 4.75          & NA                & 6.61          & 3.98          \\
            4.05            & 1.65              & 3.46          & 4.96          & NA                & 7.74          & 4.81          \\
            5.00            & 1.28              & 3.93          & 5.59          & NA                & 8.61          & 5.48          \\
            \midrule
            Avg.            & 1.56              & 2.20          & 4.62          & NA                & 6.61          & 4.37          \\
            \bottomrule
        \end{tabular}
    }
    \end{center}
    \footnotesize{\textsuperscript{\emph{a}} ICTP-LR(M)$^\ast$ could not be used to perform longer MD simulations for \ce{NaCl}-water mixtures.}\\
\end{table*}

\begin{table*}[t!]
    \caption{Backbone dihedral angles ($\phi$ and $\psi$) and free energies of representative low‑energy conformers of cationic and blocked Ala3 in explicit aqueous solution. Free energies are provided relative to the lowest-energy conformer. Backbone dihedral angles are given in degrees, while free relative energies are given in eV. Entries are shown as $\phi$/$\psi$/relative free energy. \label{tab:ala3_conformations}}
    \begin{center}
        \resizebox{\textwidth}{!}{
        \begin{tabular}{lrrrrrr}
            \toprule
                                            & \multicolumn{3}{c}{Cationic Ala3}                                                                 & \multicolumn{3}{c}{Blocked Ala3}                                                                                              \\
            Conformation                    & AMBER14SB+TIP3P       & ICTP-LR(S)        & ICTP-SR(M)        & AMBER14SB+TIP3P   & ICTP-LR(S)        & ICTP-SR(M)        \\
            \cmidrule(lr){1-1} \cmidrule(lr){2-4} \cmidrule(lr){5-7}
            Antiparallel $\beta$-sheet        & -150.5/157.0/0.040 & -159.1/162.0/0.003 & -154.8/159.8/0.021 & -149.0/159.1/0.049 & -158.4/159.1/0.000 & -153.4/156.2/0.040 \\
            Right-handed $\alpha$-helix       & -69.8/-27.4/0.059  & -72.7/-22.3/0.029  & -68.4/-25.9/0.025  & -69.1/-24.5/0.026  & -76.3/-19.4/0.011  & -67.7/-26.6/0.011  \\
            Left-handed $\alpha$-helix        & 54.7/30.2/0.096    & 64.8/31.0/0.100    & 64.1/30.2/0.131    & 53.3/30.2/0.071    & 64.1/30.2/0.083    & 61.2/30.2/0.079    \\
            PPII-type structure               & -66.2/155.5/0.000  & -68.4/146.9/0.000  & -67.0/151.9/0.000  & -67.7/154.1/0.000  & -76.3/137.5/0.005  & -63.4/144.7/0.000  \\
            \bottomrule
        \end{tabular}}
    \end{center}
\end{table*}

\begin{table*}[t!]
    \caption{$J$-coupling constants for cationic and blocked Ala3 derived from dihedral angle distributions. All $J$-coupling constants are provided in Hz. \label{tab:ala3_jcoupling}}
    \begin{center}
        \resizebox{\textwidth}{!}{
        \begin{tabular}{llllllllll}
            \toprule
                                                            & \multicolumn{4}{c}{Cationic Ala3}                                                                 & \multicolumn{5}{c}{Blocked Ala3}                                                                                              \\
            Coupling                                        & Experiment\cite{Graf2007}  & AMBER14SB+TIP3P\textsuperscript{\emph{a}}   & ICTP-LR(S)        & ICTP-SR(M)        & AMBER14SB+TIP3P   & ANI-2x\cite{Rosenberger2021}  & MACE-OFF24(M)\cite{Kovacs2025}    & ICTP-LR(S)        & ICTP-SR(M)        \\
            \cmidrule(lr){1-1} \cmidrule(lr){2-5} \cmidrule(lr){6-10}
            $^3J(\mathrm{H}_\mathrm{N},\mathrm{H}_\alpha)$  & 5.68 $\pm$ 0.11                       & 6.07 (5.93 $\pm$ 2.09)            & 7.17 $\pm$ 2.16   & 6.60 $\pm$ 2.27   & 6.01 $\pm$ 2.06   & 7.50 $\pm$ 2.02               & 5.70 $\pm$ 0.43                   & 7.40 $\pm$ 2.06   & 6.60 $\pm$ 2.32   \\
            $^3J(\mathrm{H}_\mathrm{N},\mathrm{C}^\prime)$  & 1.13 $\pm$ 0.08                       & 1.13 (1.76 $\pm$ 1.09)            & 2.19 $\pm$ 1.44   & 2.05 $\pm$ 1.31   & 1.57 $\pm$ 0.94   & 1.80 $\pm$ 1.88               & 1.37 $\pm$ 0.30                   & 2.19 $\pm$ 1.44   & 1.72 $\pm$ 1.10   \\
            $^3J(\mathrm{H}_\alpha,\mathrm{C}^\prime)$      & 1.84 $\pm$ 0.13                       & 1.70 (1.74 $\pm$ 0.93)            & 2.11 $\pm$ 1.02   & 1.86 $\pm$ 0.76   & 1.83 $\pm$ 1.13   & 2.20 $\pm$ 0.91               & --                                & 2.10 $\pm$ 0.83   & 1.98 $\pm$ 1.09   \\
            $^3J(\mathrm{C}^\prime,\mathrm{C}^\prime)$      & 0.25 $\pm$ 0.10                       & 0.79 (0.83 $\pm$ 0.70)            & 1.27 $\pm$ 0.89   & 1.09 $\pm$ 0.84   & 0.73 $\pm$ 0.58   & 1.40 $\pm$ 0.88               & --                                & 1.29 $\pm$ 0.88   & 0.90 $\pm$ 0.68   \\
            $^3J(\mathrm{H}_\mathrm{N},\mathrm{C}_\beta)$   & 2.39 $\pm$ 0.09                       & 1.87 (1.85 $\pm$ 0.81)            & 1.32 $\pm$ 0.88   & 1.56 $\pm$ 0.88   & 1.91 $\pm$ 0.74   & 1.20 $\pm$ 0.89               & 1.64 $\pm$ 0.18                   & 1.28 $\pm$ 0.85   & 1.68 $\pm$ 0.82   \\
            $^1J(\mathrm{N},\mathrm{C}_\alpha)$             & 11.34 $\pm$ 0.07                      & 11.41 (9.82 $\pm$ 0.79)           & 10.30 $\pm$ 1.01  & 10.11 $\pm$ 0.96  & 9.71 $\pm$ 0.66   & 10.50 $\pm$ 0.90              & --                                & 10.32 $\pm$ 1.01  & 9.89 $\pm$ 0.82   \\
            \bottomrule
        \end{tabular}}
    \end{center}
    \footnotesize{\textsuperscript{\emph{a}} Values in parentheses were obtained using the flexible TIP3P water model and reweighted dihedral angle probability densities from metadynamics simulations, ensuring consistency with the other results presented in this work. Values not in parentheses are from Ref.~\citenum{Zhang2020}.}
\end{table*}

\clearpage

\clearpage

\begin{table*}[t!]
	\caption{Overview of the training and test datasets used in this work. Most datasets are taken from SPICE-v2.\cite{Eastman2023b, Eastman2024} For all other datasets, the corresponding references are explicitly provided. \label{tab:datasets}}
	\begin{center}
    \resizebox{\textwidth}{!}{
	\begin{tabular}{lllll}
	\toprule 
	Dataset		    		                & $N_\mathrm{train+valid}$  & $N_\mathrm{test}$     & $N_\mathrm{at}$       & Chemical elements                                 \\
	\midrule
    Dipeptides                                  & 32157                     & 1693                  & 26--60                & H, C, N, O, S                                     \\
    Solvated Amino Acids                        & 1235                      & 65                    & 79--96                & H, C, N, O, S                                     \\
    DES370K Dimers                              & 328390                    & 17286                 & 2--34                 & H, Li, C, N, O, F, Na, Mg, P, S, Cl, K, Ca, Br, I \\
    DES370K Monomers                            & 17765                     & 935                   & 3--22                 & H, C, N, O, F, P, S, Cl, Br, I                    \\
	PubChem                                     & 1327459                   & 69878                 & 3--50                 & H, B, C, N, O, F, Si, P, S, Cl, Br, I             \\
    Solvated PubChem                            & 13230                     & 697                   & 63--110               & H, C, N, O, F, P, S, Cl, Br, I                    \\
    Amino Acid Ligand Pairs                     & 184316                    & 9702                  & 24--72                & H, C, N, O, F, P, S, Cl, Br, I                    \\
    Ion Pairs                                   & 1356                      & 72                    & 2                     & Li, F, Na, Cl, K, Br, I                           \\
    Water Clusters\textsuperscript{\emph{a}}    & 2546                      & 135                   & 3--150                & H, O                                              \\
    QMugs\cite{Isert2022,Kovacs2025}            & 2746                      & 146                   & 50--90                & H, C, N, O, F, P, S, Cl, Br, I                    \\
    \ce{NaCl}-Water Clusters                    & 1036                      & 56                    & 60--150               & H, O, Na, Cl                                      \\
    \midrule
    \multicolumn{5}{l}{Test-only datasets}                                                                                                                                     \\
    \midrule
    Small Ligands\cite{Eastman2024}             & --                        & 1996                  & 40--50                & H, B, C, N, O, F, P, S, Cl, Br, I                 \\
    Large Ligands\cite{Eastman2024}             & --                        & 1994                  & 70--80                & H, C, N, O, F, P, S, Cl, Br, I                    \\
    Pentapeptides\cite{Eastman2024}             & --                        & 2000                  & 68--110               & H, C, N, O, S                                     \\
    TorsionNet-500\cite{Rai2022,Kovacs2025}     & --                        & 12000                 & 13--37                & H, C, N, O, F, S, Cl                              \\
    Biaryl\cite{Lahey2020,Kovacs2025}           & --                        & 2112                  & 17--28                & H, C, N, O, S                                     \\
	\bottomrule 
	\end{tabular}}
	\end{center}
    \footnotesize{\textsuperscript{\emph{a}} We combine water clusters from the SPICE-v2 dataset\cite{Eastman2024} with those used in Ref.~\citenum{Kovacs2025}.}
\end{table*}

\end{appendices}

\end{document}